\documentclass[english, a4paper]{article}

\usepackage[margin=1.65in]{geometry}
\usepackage[font=small, labelfont=bf, margin=0.35in]{caption}

\usepackage{titling}
\usepackage{authblk}
\usepackage{bm}

\usepackage{babel}
\usepackage[utf8]{inputenc}
\usepackage[pdftex]{hyperref}

\usepackage{xcolor}

\usepackage{amsmath}
\usepackage{amssymb}
\usepackage{amsthm}

\usepackage{cite}

\usepackage{pdflscape}

\usepackage{enumitem}

\usepackage{multirow}
\usepackage{siunitx}
\usepackage{array}

\makeatletter
\newcommand{\thickhline}{%
  \noalign {\ifnum 0=`}\fi \hrule height 1pt
  \futurelet \reserved@a \@xhline
}
\newcolumntype{"}{@{\hskip\tabcolsep\vrule width 1pt\hskip\tabcolsep}}
\makeatother

\newcommand{\thickcline}[1]{%
  \@thickcline #1\@nil%
}

\def\@thickcline#1-#2\@nil{%
  \omit
  \@multicnt#1%
  \advance\@multispan\m@ne
  \ifnum\@multicnt=\@ne\@firstofone{&\omit}\fi
  \@multicnt#2%
  \advance\@multicnt-#1%
  \advance\@multispan\@ne
  \leaders\hrule\@height1pt\hfill
  \cr
  \noalign{\vskip-1pt}%
}

\usepackage[labelfont=bf, figurename=Fig., tablename=Tab.]{caption}

\allowdisplaybreaks

\newcommand{\refeq}[1]{(\ref{eq:#1})}

\newcommand{\ket}[1]{\left|\, #1 \, \right\rangle}

\newcommand{\imag}{\mathrm{i}}
\newcommand{\ceil}[1]{\left\lceil #1 \right\rceil}
\newcommand{\floor}[1]{\left\lfloor #1 \right\rfloor}
\newcommand{\round}[1]{\left\lfloor #1 \right\rceil}
\newcommand{\norm}[1]{\left\lVert #1 \right\rVert}

\newcommand{\vect}[1]{\bm{#1}}

\newcommand{\cexp}{w}

\title{A high-level comparison of state-of-the-art quantum algorithms for breaking asymmetric cryptography}
\author[1,2]{\href{mailto:ekera@kth.se}{Martin Ekerå}}
\author[1,2]{\href{mailto:jgartner@kth.se}{Joel Gärtner}}
\affil[1]{\small KTH Royal Institute of Technology, Stockholm, Sweden}
\affil[2]{\small Swedish NCSA, Swedish Armed Forces, Stockholm, Sweden}

\predate{\begin{center}}
\postdate{\end{center}\vspace{-3mm}}

\begin{document}
\maketitle

\begin{abstract}
  We provide a high-level cost comparison between Regev's quantum algo\-rithm with Ekerå--Gärtner's extensions on the one hand, and existing state-of-the-art quantum algorithms for factoring and computing discrete logarithms on the other.
  This when targeting cryptographically relevant problem instances, and when accounting for the space-saving optimizations of Ragavan and Vaikuntanathan that apply to Regev's algorithm, and optimizations such as windowing that apply to the existing algorithms.

  Our conclusion is that Regev's algorithm without the space-saving optimizations may achieve a per-run advantage, but not an overall advantage, if non-computational quantum memory is cheap.
  Regev's algorithm with the space-saving optimizations does not achieve an advantage, since it uses more computational memory, whilst also performing more work, per run and overall, compared to the existing state-of-the-art algorithms.
  As such, further optimizations are required for it to achieve an advantage for cryptographically relevant problem instances.
\end{abstract}

\section{Introduction}
\label{sect:introduction}
In August of~2023, Regev~\cite{regev23} introduced a quantum factoring algorithm that may be perceived as a $d$-dimensional variation of Shor's factoring algorithm~\cite{shor94, shor97}.

To factor an $n$-bit integer~$N$, Regev raises the squares of the first~$d = \ceil{\sqrt{n}\,}$ primes to short exponents.
By using binary tree-based arithmetic, and square-and-multiply--based exponentiation, Regev achieves a circuit size reduction by a factor $\Theta(\sqrt{n})$ compared to the other state-of-the-art variations of Shor's algorithms that are in the literature.
This reduction comes at the expense of having to perform $d + 4$ runs, however, and at the expense of using $O(n^{3/2})$ qubits of space in each run.
Regev's algorithm furthermore relies on a heuristic number-theoretic assumption.

In October of~2023, Ragavan and Vaikuntanathan~\cite{rv23} reduced the space requirements to~$\tilde O(n)$ qubits by using Fibonacci-based exponentiation.
A few months later, in February of~2024, Ragavan and Vaikuntanathan~\cite{rv23v3} improved the constants in their analysis and assigned a new title to their pre-print.
Very recently, in April and May of 2024, Ragavan~\cite{ragavan24} introduced further optimizations, and generalized the Fibonacci-based exponentiation so as to allow for tradeoffs between the circuit size and the space usage of the circuit.

In November of~2023, Ekerå and Gärtner~\cite{eg23-dlog} extended Regev's factoring algo\-rithm to algorithms for computing discrete logarithms and orders quantumly in groups for which there exist a notion of~\emph{small} elements.
Furthermore, Ekerå and Gärtner explained how to factor integers completely via order finding.
The resulting factoring algorithm is slightly more efficient per run compared to Regev's original algorithm.
Note, however, that Ekerå and Gärtner's algorithms rely on a slightly stronger heuristic number-theoretic assumption.

In April of~2024, Pilatte~\cite{pilatte24} proved a version of this assumption by using tools from analytic number theory.
This allows Ekerå--Gärtner's extensions of Regev's algorithm to be used to unconditionally factor and compute discrete logarithms, although this requires selecting parameters in a manner that may not be preferable in practice.

\subsection{Our contributions}
Arguably, quantum algorithms for factoring integers and computing discrete logarithms are of interest primarily because such algorithms may be used to break currently widely deployed asymmetric cryptography --- including but not limited to Rivest--Shamir--Adleman (RSA)~\cite{rsa, nist-sp-800-56b}, Diffie--Hellman (DH)~\cite{diffie-hellman, nist-sp-800-56a} and DSA~\cite{fips186-dss, fips186-5-dss}, and their elliptic curve counterparts EC-DH~\cite{nist-sp-800-56a} and EC-DSA~\cite{fips186-5-dss}.

Although the circuit for Regev's algorithm is asymptotically smaller than the circuits for the other state-of-the-art variations of Shor's algorithms, the asymptotic advantage only kicks in for sufficiently large problem instances.
Furthermore, the circuit for Regev's algorithm uses more space, and asymptotically more runs of the circuit are required to achieve a high success probability.
Hence, it is not clear that Regev's algorithm has an advantage in practice for cryptographically relevant problem instances, and the same holds true for Ekerå--Gärtner's extensions.

To begin to resolve this situation, in this work, we provide a coarse high-level comparison between Regev's algo\-rithm and Ekerå--Gärtner's extensions of it on the one hand, and some of the other state-of-the-art quantum algorithms~\cite{ekera-hastad, ekera-pp, ekera-short-success, ekera-general, ekera-revisiting} on the other hand.
In particular, we consider factoring RSA integers to break RSA, and computing discrete logarithms in subgroups to~$\mathbb Z_N^*$ for~$N$ a large prime to break DH and DSA.
We do not consider EC-DH and EC-DSA since these schemes work in elliptic curve groups that do not admit a trivial notion of small group elements.

For factoring RSA integers and computing short discrete logarithms, we compare Ekerå--Gärtner's extensions of Regev's algorithm to Ekerå--Håstad's variation~\cite{ekera-hastad, ekera-pp, ekera-short-success} of Shor's algorithm.
For computing general discrete logarithms, we instead compare them to Ekerå's variation~\cite{ekera-revisiting} of Shor's algorithm.

We account for Ragavan's generalization~\cite{ragavan24} of the space-saving optimization of Ragavan and Vaikuntanathan~\cite{rv23v3} that apply to Regev's algorithm and to Ekerå--Gärtner's extensions, and for optimization such as windowing~\cite{gidney-ekera, gidney-window, vmi05, van-meter} that apply to most other variations of Shor's algorithms, including in particular Ekerå--Håstad's and Ekerå's variations.

We consider a range of problem instance sizes from currently widely deployed to more conservative.
Since all of the aforementioned schemes are being phased out, we do not foresee that larger key sizes than what we cost in this paper will be widely deployed in the future.

\subsection{Methodology}
\label{sect:methodology}
Regev's algorithm~\cite{regev23} and Ekerå--Gärtner's extensions~\cite{eg23-dlog} use more space, and require more runs, than Ekerå--Håstad's~\cite{ekera-hastad, ekera-pp, ekera-short-success} and Ekerå's~\cite{ekera-revisiting} variations of Shor's algorithm --- hereinafter referred to as the existing algorithms.
This is true also when accounting for space-saving optimizations of Ragavan and Vaikuntanathan~\cite{rv23v3}, and Ragavan's generalization~\cite{ragavan24}.
It follows that for Regev's algorithm and Ekerå--Gärtner's extensions to achieve an advantage per run in practice over the existing algorithms, they need to perform less work per run.

To compare the amount of work performed per run, we follow Ragavan and Vaikuntanathan~\cite{rv23v3, ragavan24}, and assume that we have a quantum circuit that maps
\begin{align}
  \label{eq:quantum-multiplication}
  \ket{u, v, t, 0^S}
  \rightarrow
  \ket{u, v, (t + uv) \text{ mod } N, 0^S}
\end{align}
for~$S$ the number of ancilla qubits, and for $u, v, t \in [0, N) \cap \mathbb Z$, where we furthermore require $u$ and $v$ to be invertible modulo~$N$.
All of the aforementioned quantum algorithms can be implemented by calling this circuit as a subroutine.
Furthermore, for the parameterizations of the aforementioned algorithms that we consider, all other work performed in a run is negligible compared to the calls to this circuit.
Hence, we may roughly compare the amount of work that the algorithms perform by counting the number of calls that they make to this multiplication circuit.

For RSA, for~$N$ an RSA integer, we count the number of circuit calls made in a run of Ekerå--Håstad's variation of Shor's algorithm, and of Regev's algorithm or of Ekerå--Gärtner's extension to factoring via order finding, respectively.

For DH and DSA, for~$N$ a large prime, we similarly count the number of circuit calls made in a run of Ekerå--Håstad's or of Ekerå's variation of Shor's algorithm, and of Ekerå--Gärtner's extension of Regev's algorithm, respectively.

Furthermore, we state the estimated number of runs required to solve the problem instance with a given high success probability, and count the overall number of circuit calls required across all of these runs.

We consider~$N$ of bit length $n \in \{ 2048, \, 3072, \, 4096, \, 6144, \, 8192 \}$.
Our rationale for this choice is that $n \in \{ 2048, \, 3072, \, 4096 \}$ are arguably the most common choices in current commercial cryptographic applications, whereas $n \in \{ 6144, \, 8192 \}$ are more conservative choices.
Note also that already for quite some time now, there has been a gradual shift away from RSA, DH and DSA --- initially to EC-DH and EC-DSA, and more recently to post-quantum secure cryptography.
Hence, we do not expect~$N$ of larger bit lengths to be widely deployed in the future.
Rather, we expect the use of RSA, DH and DSA to continue to decrease, and eventually cease.

For DSA, that uses Schnorr groups, we consider subgroups of sizes selected based on the strength level estimates used by NIST, as given in~\cite[Tab.~25--26 in App.~D on p.~133]{nist-sp-800-56a} and~\cite[Sect.~7.5 on p.~126]{fips-140-2-guidance}.
For DH, we consider both safe-prime groups with short exponents and Schnorr groups.
We select the logarithm length and subgroup size, respectively, based on the strength level estimates used by NIST.
Note that for DH, the security typically depends on the short discrete logarithm problem in practice in protocols such as TLS~\cite{rfc7919} and IKE~\cite{rfc3526}.

\subsection{Overview}
In what follows, we first review Ekerå--Håstad's and Ekerå's variations of Shor's algorithm in Sect.~\ref{sect:existing-variations}.
We then review Regev's algorithm and Ekerå--Gärtner's extensions of it in Sect.~\ref{sect:regev-extensions}.
We describe the outcome of our cost comparison in Sect.~\ref{sect:cost-comparisons-results}, and summarize and conclude the paper in Sect.~\ref{sect:summary-conclusion}.
We provide detailed tabulated results in App.~\ref{app:tables}.

\subsection{Notation}
In what follows, we denoted by $\log$ the base-two logarithm.
We use~$\ceil{u}$, $\floor{u}$ and~$\round{u}$ to denote~$u$ rounded up, down and to the closest integer, respectively.
We use the convention that empty products evaluate to one.

\section{Existing variations of Shor's algorithm}
\label{sect:existing-variations}
In this section, we briefly review Ekerå--Håstad's~\cite{ekera-hastad, ekera-pp, ekera-short-success} and Ekerå's~\cite{ekera-revisiting} variations of Shor's algorithm~\cite{shor94, shor97}, so as to introduce relevant notation, and so as to describe implementation details relevant to our cost comparison.

\subsection{Ekerå--Håstad's variation of Shor's algorithm}
Ekerå--Håstad's variation~\cite{ekera-hastad, ekera-pp, ekera-short-success} of Shor's algorithm~\cite{shor94, shor97} (EHS) efficiently computes short discrete logarithms quantumly in cyclic groups of unknown order.
It furthermore efficiently breaks RSA via a reduction from the RSA integer factoring problem (IFP) to the short discrete logarithm problem (DLP).

The algorithm induces the state
\begin{align*}
  &\frac{1}{\sqrt{2^{m + 2 \ell}}}
  \sum_{a \, = \, 0}^{2^{m+\ell} - 1}
  \sum_{b \, = \, 0}^{2^{\ell} - 1}
  \ket{a, b, g^{a} x^{-b}}
  \xrightarrow{\text{QFT}} \\
  &\quad\quad \frac{1}{2^{m + 2 \ell}}
  \sum_{a, \, j \, = \, 0}^{2^{m+\ell} - 1}
  \sum_{b, \, k \, = \, 0}^{2^{\ell} - 1}
  \exp\left[ \frac{2 \pi \imag}{2^{m+\ell}} (aj + 2^m bk) \right]
  \ket{j, k, g^{a} x^{-b}}
\end{align*}
where $g \in \mathbb Z_N^*$ in the cases we consider in this paper, for~$N$ an RSA integer or a prime, $x = g^d$ for~$d$ a short $m$-bit logarithm, and $\ell \sim m / s$ for~$s$ a tradeoff factor.

By increasing~$s$, the amount of work that needs to be performed in each run of the algorithm is reduced, at the expense of more runs being required to solve for~$d$ in the classical post-processing.
Each run yields approximately $\ell$~bits of information on~$d$, so the idea is to perform $n \ge s$ runs and to jointly solve the~$n$ outputs for~$d$.
In practice, this is achieved by using classical lattice-based post-processing.

For an analysis of how many runs~$n$ are required for a given tradeoff factor~$s$ and logarithm length~$m$ when breaking RSA, or when solving short discrete logarithms in safe-prime groups, see~\cite{ekera-pp}.
For an analysis of how much work is required when picking $\ell = m - \Delta$ for some small $\Delta$ and solving in a single run, see~\cite{ekera-short-success}.

\subsubsection{Implementing the algorithm}
\label{sect:implementing-ekera-hastad}
Although there are more elaborate ways to implement EHS in an optimized fashion --- see e.g.~\cite{gidney-ekera} --- a straightforward way to implement the algorithm is to classically pre-compute powers of the group elements that are to be exponentiated, and to compose them quantumly under the group operation --- in this case multiplication modulo~$N$ --- conditioned on a control qubit.

For this purpose, we essentially need a circuit that maps
\begin{align*}
  \ket{c_i, u} \rightarrow \ket{c_i, u \cdot v_i^{c_i} \text{ mod } N}
\end{align*}
for $v_i$ the classically pre-computed value, and~$c_i \in \{0, 1\}$ the control qubit.

Such a circuit may be constructed from the circuit in~\refeq{quantum-multiplication} by loading $v_i^{c_i}$ --- i.e.\ either~$1$ or~$v_i$ conditioned on~$c_i$ --- into a quantum register, and then taking
\begin{align*}
  \ket{c_i, u, v_i^{c_i}, 0, 0^S}
  \rightarrow
  \ket{c_i, u, v_i^{c_i}, u \cdot v_i^{c_i} \text{ mod } N, 0^S}
\end{align*}
after which we unload $v_i^{c_i}$, load $-v_i^{-c_i} \text{ mod } N = -(v_i^{-1})^{\, c_i} \text{ mod } N$, and take
\begin{align*}
  \ket{c_i, u, -v_i^{-c_i} \text{ mod } N, u \cdot v_i^{c_i} \text{ mod } N, 0^S}
  \rightarrow
  \ket{c_i, 0, -v_i^{-c_i} \text{ mod } N, u \cdot v_i^{c_i} \text{ mod } N, 0^S}
\end{align*}
after which we unload $-v_i^{-c_i} \text{ mod } N$ to obtain
\begin{align*}
  \ket{c_i, u \cdot v_i^{c_i} \text{ mod } N, 0, 0, 0^S}
\end{align*}
as desired.
A few intermediary swaps are required.
We may then proceed to process the next element and control qubit, recursively.
As we run through $i \in [0, n_e) \cap \mathbb Z$, for~$n_e$ the total exponent length, we make a total of~$2 n_e$ calls to the circuit.

Following the exponentiation, two quantum Fourier transforms (QFTs) are applied to the control registers, and the result~$(j, k)$ read out.
To save space in practical implementations, the two QFTs may be interleaved with the exponentiation, and the control qubit recycled~\cite{parker-plenio, mosca-ekert}.
A single qubit then suffices to implement the control registers for the exponentiation.
Note also that small phase shifts may be dropped by using Coppersmith's~\cite{coppersmith2002} approximate QFT.

As a further optimization, we may use windowing~\cite{gidney-ekera, gidney-window, vmi05, van-meter}, and process the control qubits in a window of some given size~$\cexp$.
That is to say, instead of loading $v_i^{c_i}$ and $-v_i^{-c_i} \text{ mod } N$, we may use a quantum lookup table to load
\begin{align*}
  \prod_{j \, = \, 0}^{\cexp - 1} v_{i+j}^{c_{i+j}} \text{ mod } N
  \quad\quad \text{ and } \quad\quad
  -\prod_{j \, = \, 0}^{\cexp - 1} v_{i+j}^{-c_{i+j}} \text{ mod } N,
\end{align*}
respectively, reducing the number of calls to the circuit to $2 \cdot \lceil n_e / \cexp \rceil$.
As explained above, the QFT may be interleaved with the exponentiation, and the control qubits recycled, in which case~$\cexp$ qubits suffice to implement the control registers.

For each windowed multiplication, a table of~$2^{\cexp}$ values is first classically pre-computed, after which a quantum lookup into this table is performed.
The classical pre-computation is not limiting, as we envisage quantum computations to be significantly more expensive than classical computations.
However, the quantum lookup is not free.
The cost of the lookup bounds how large~$\cexp$ can reasonably be selected if we are to be able to neglect the lookup cost when comparing it to the cost of performing a large multiplication modulo~$N$.

As explained in~\cite{gidney-ekera, gidney-window}, it is possible to go further and to combine windowing over the control qubits in the exponent with windowing over control qubits inside the multiplication circuit, and to optimize the multiplication circuit in various ways.

More specifically, in~\cite{gidney-ekera} for RSA-2048, these two levels of windowing are both of size~$5$, resulting in the algorithm performing lookups in a table of size~$2^{10}$.
The window over the exponent decreases the number of multiplications by a factor of~$5$, and the window inside the multiplication circuit reduces the cost of this circuit by a factor of~$5$.
In total, the use of windowing hence essentially decreases the cost of the algorithm as a whole by a factor of~$25$.

For our cost estimates, we consider a black-box multiplication circuit, which we can not optimize, and we therefore can not use two levels of windowing.
Instead, for simplicity, and for the same lookup cost, we use a single window of size~$\cexp = 10$ over the exponent to decrease the number of multiplications by a factor of~$10$.
Note that this under\-estimates the efficiency gain from windowing in practice, as a larger gain would be possible if using a non-black-box multiplication circuit.

Note furthermore that selecting a larger window size could be beneficial for larger problem sizes.
In practice, it would be reasonable to for example select $\cexp = \Theta(\log \log N)$, but for simplicity we fix $w = 10$ for now in our cost comparisons.
This when having a circuit for schoolbook multiplication in mind, as this seems to be the optimal choice~\cite{gidney-window} for the problem sizes we consider in this paper.

\subsection{Ekerå's variation of Shor's algorithm}
Ekerå~\cite{ekera-revisiting} has modified Shor's algorithm for computing general discrete logarithms so as to obtain an algorithm (ES) that efficiently solves general discrete logarithms in cyclic groups of known order.
The resulting algorithm induces the state
\begin{align*}
  &\frac{1}{\sqrt{2^{m + \varsigma + \ell}}}
  \sum_{a \, = \, 0}^{2^{m+\varsigma} - 1}
  \sum_{b \, = \, 0}^{2^{\ell} - 1}
  \ket{a, b, g^{a} x^{-b}}
  \xrightarrow{\text{QFT}} \\
  &\quad\quad \frac{1}{2^{m+\varsigma+\ell}}
  \sum_{a, \, j \, = \, 0}^{2^{m+\varsigma} - 1}
  \sum_{b, \, k \, = \, 0}^{2^{\ell} - 1}
  \exp\left[ \frac{2 \pi \imag}{2^{m+\ell}} (aj + 2^{m+\ell-\varsigma} bk)\right]
  \ket{j, k, g^{a} x^{-b}}
\end{align*}
where $g \in \mathbb Z_N^*$ in the cases we consider in this paper, for~$N$ a prime, $x = g^d$ for $d \in [0, r) \cap \mathbb Z$, $m$ the bit length of the order~$r$ of~$g$, and $\ell \sim m / s$ for~$s$ a tradeoff factor and~$\varsigma$ a parameter to suppress noise.
As for EHS, each run of ES yields approximately $\ell$~bits of information on~$d$.
The idea is to perform $n \ge s$ runs and to jointly post-process the outputs given~$r$ using lattice-based post-processing.

As is explained in~\cite{ekera-revisiting}, a problem that arises when making these modifications is that noise appears in the distribution.
When solving in a single run with $s = 1$, this is not a problem since we may overcome the noise by searching a bit in the post-processing.
When picking $s > 1$ and solving in $n \ge s$ runs, the parameter~$\varsigma$ may be increased slightly above zero to suppress the noise.
Hence, we pick $\varsigma = 0$ when solving in a single run and some small~$\varsigma$ when solving in many runs.

Note that besides enabling tradeoffs, the modifications made to the algorithm enable qubit recycling by ensuring that the control qubits are separable.
This is not the case in Shor's original algorithm where the control registers run over $[0, N-1) \cap \mathbb Z$.
The modifications furthermore make the complexity of the algorithm depend on the bit length~$m$ of~$r$, rather than on~$N$ as in Shor's original algorithm.

\subsubsection{Implementing the algorithm}
ES may be implemented in the same way as EHS, using control qubit recycling, approximate QFTs, windowing, etc.
For further details, see Sect.~\ref{sect:implementing-ekera-hastad}.
The total exponent length is $n_e = m + \varsigma + \ell$, and there are $2 \ceil{n_e / \cexp}$ calls to the multiplication circuit, for~$\cexp$ the window size.

\section{Regev's algorithm and its extensions}
\label{sect:regev-extensions}
In Regev's algorithm~\cite{regev23}, and Ekerå--Gärtner's extensions~\cite{eg23-dlog} thereof, a state proportional to
\begin{align}
  \label{eq:regev-quantum-state}
  \sum_{\vect z \, \in \, \{ -D/2, \ldots, D/2 - 1 \}^{d+k}}
  \hspace{-6mm}
  \rho_R(\vect z)
  \ket{z_1, \ldots, z_{d+k},
  \prod_{i \, = \, 1}^d a_i^{z_i + D/2}
  \prod_{i \, = \, 1}^k x_i^{z_{d + i} + D/2}
  \text{ mod } N}
\end{align}
is first induced, where~$d$ and~$k$ are parameters, $\vect z = (z_1, \ldots, z_{d+k})$, and the~$d$ integers $a_1, \ldots, a_d$ are small whereas the $k \ge 0$ integers $x_1, \ldots, x_k$ may be arbitrarily selected.
Furthermore, $R = 2^{C \sqrt{n}}$ for $C > 0$ a constant, $D = 2^{\lceil \log(2 \sqrt{d} R) \rceil}$, and
\begin{align*}
  \rho_R(\vect z) = \exp \left( -\frac{\pi}{R^2} \, |\vect z|^2 \right)
\end{align*}
is a Gaussian function.
QFTs are then applied to the first~$d+k$ control registers, and the registers read out, yielding a vector $\vect w = (w_1, \ldots, w_{d+k}) / D \in [0, 1)^d$.

The quantum circuit is run $m \ge d + 4$ times.
The resulting vectors $\vect w_1, \ldots, \vect w_m$ are then post-processed classically using lattice-based techniques to recover the solution with probability $\ge 1/4$ under a heuristic assumption.

\subsection{Implementing the algorithm}
In this section, we first explain how to implement Regev's original algorithm for which~$k = 0$, and then consider the case $k > 0$ as it is relevant to some of Ekerå and Gärtner's extensions of Regev's algorithm.

\subsubsection{Regev's original arithmetic}
To compute the product over $a_1, \ldots, a_d$ in the last register in~\refeq{regev-quantum-state}, Regev originally proposed to process the first~$d$ registers $z_1, \ldots, z_d$ bit by bit.
Let $D = 2^l$, and
\begin{align*}
  z_i + D/2 = \sum_{j \, = \, 0}^{l-1} 2^j z_{i,j}
  \quad
  \text{ for }
  \quad
  z_{i,j} \in \{ 0, 1 \}.
\end{align*}

Then the product in the last register may be re-written as
\begin{align*}
  \prod_{i \, = \, 1}^d a_i^{z_i+D/2} \text{ mod } N
  =
  \prod_{j \, = \, 0}^{l-1}
  \left(\, \prod_{i \, = \, 1}^d a_i^{z_{i, j}} \right)^{2^j}
  \hspace{-2mm} \text{ mod } N
  =
  \prod_{j \, = \, 0}^{l-1}
  c_j^{\, 2^j} \text{ mod } N
\end{align*}
where we denote the inner product over~$i$ by~$c_j$.

In Regev's algorithm, the~$a_i$ are small integers, so~$c_j$ is the product of small integers~$a_i$ raised to powers $z_{i,j} \in \{0, 1\}$.
For the values of~$d$ considered by Regev, the product~$c_j$ is furthermore guaranteed to be significantly smaller than~$N$, ensuring that~$c_j$ can be computed quite efficiently by using a binary tree-based approach, as described by Regev.
When estimating the cost of Regev's algorithm, we therefore ignore the cost of computing and uncomputing~$c_j$.

To compute the full product given the~$c_j$, Regev proposes to use the square-and-multiply algorithm:
The first step is to compute $c_{l-1}$ to an intermediate register.
Next, the square modulo~$N$ of this register is computed to another intermediate register and multiplied with~$c_{l-2}$ modulo~$N$.
The algorithm continues this process recursively until~$c_{0}$ is multiplied into the last intermediary register, which then contains the desired product.
Finally, the result is copied out to a separate register, and the whole circuit run in reverse to uncompute the intermediary registers.

In total the algorithm thus performs $2(l-1)$ squarings modulo~$N$, and $2(l-1)$ multiplications with~$c_j$ modulo~$N$ for different $j$.
A squaring or multiplication modulo~$N$ can be performed in a single call to the circuit in~\refeq{quantum-multiplication}.
The full algorithm can thus be implemented by calling this circuit $4(l-1)$ times.

Unfortunately, however, the algorithm also requires $l + 1= O(\log D) = O(\sqrt{n})$ registers, each of length $n$~qubits, to hold the intermediate values generated by the squaring operations and the final result.
This is necessary since squaring modulo~$N$ is not a reversible operation; it can not be efficiently performed in place.

Note also that windowing can not be applied to Regev's algorithm, since both operands passed to the multiplication circuit are quantum in his algorithm, whereas one of the operands is a conditionally loaded classical constant in most other variations of Shor's algorithms.
Windowing is used to load the constant operand in an efficient manner:
It critically relies on the values in the lookup table being classical constants that can be pre-computed.
Only the lookup index is quantum.

\subsubsection{Ragavan--Vaikuntanathan's space-saving optimizations}
\label{sect:rv-optimizations}
To circumvent the reversibility issue in Regev's original arithmetic, and the space increase to which it gives rise, Ragavan and Vaikuntanathan~\cite{rv23, rv23v3} have proposed to compute the product
\begin{align}\label{eq:product-ai}
    \prod_{i \, = \, 1}^d a_i^{z_i+D/2} \text{ mod } N
\end{align}
in a different way, that does not require non-invertible squarings modulo~$N$ to be performed.
They accomplish this feat by decomposing the product with Fibonacci numbers in the exponent instead of powers of two, by writing
\[
    z_i + D/2 = \sum_{j \, = \, 1}^K z_{i,j} F_j
    \quad
    \text{ for }
    \quad
    z_{i, j} \in \{0, 1\},
\]
where $F_j$ is the $j$:th Fibonacci number, and~$K$ is the greatest integer such that ${F_K \le D}$.
In a recent follow-up work, Ragavan~\cite{ragavan24} generalizes this idea by defining generalized Fibonacci numbers via the recurrence
\begin{align*}
  G_j^{(r)} = rG_{j-1}^{(r)} + G_{j-2}^{(r)},
  \quad\quad
  G_1^{(r)} = G_0^{(r)} = 1,
\end{align*}
for~$r$ some positive integer.
Ragavan decomposes the product~\refeq{product-ai} with these generalized Fibonacci numbers in the exponent, by writing
\[
    z_i + D/2 = \sum_{j \, = \, 1}^{K^{(r)}} z_{i,j} G_j^{(r)}
    \quad
    \text{ for }
    \quad
    z_{i, j} \in \{0, 1, \ldots, r\},
\]
where~$K^{(r)}$ is the greatest integer such that $G_{K^{(r)}}^{(r)} \le D$.

Note that the generalized Fibonacci numbers reduce to the ordinary Fibonacci numbers for $r = 1$, so we describe only Ragavan's generalization in what follows.

Note furthermore that re-writing a number in the generalized Fibonacci basis can be accomplished efficiently and essentially in-place, as explained by Ragavan~\cite{ragavan24}.
The cost is negligible compared to the cost of performing large multiplications modulo~$N$, so we neglect it in our cost estimates.

The product~\refeq{product-ai} is rewritten as
\begin{align}
    \label{eq:compute}
    \prod_{i \, = \, 1}^d a_i^{z_i+D/2} \text{ mod } N =
    \prod_{j \, = \, 1}^{K^{(r)}}
    \left( \: \prod_{i \, = \, 1}^d a_i^{z_{i, j}} \right)^{G_j^{(r)}} \hspace{-5mm} \text{ mod } N =
    \prod_{j \, = \, 1}^{K^{(r)}}
    c_j^{G_j^{(r)}} \text{ mod } N
\end{align}
where, for each~$j$, we denote the inner product over~$i$ by~$c_j$.

As~$c_j$ is the product of~$d$ small integers~$a_i$ raised to powers $z_{i,j} \in {\{0, 1, \ldots, r\}}$, for relatively small~$d$ and~$r$, it can be computed quite efficiently by generalizing Regev's binary tree-based arithmetic.
In our cost estimates, we consider the cost of computing~$c_j$ to be negligible and therefore do not account for it.

Note however, that this cost model is only reasonable when
\[
    \prod_{i \, = \, 1}^d a_i^r < N
\]
as large multiplications modulo~$N$ are otherwise required to compute~$c_j$.
This limits which combinations of~$d$ and~$r$ we consider in our cost estimates.

To compute the product~\refeq{product-ai}, Ragavan introduces an additional parameter~$s$ that is a power of two such that~$r$ is a multiple of~$s$.
By using $4 \log s$ additional intermediary $n$-bit registers, Ragavan computes the product with $f(r, s) \cdot K^{(r)}$ large multiplications modulo~$N$, where
\begin{align}
  \label{eq:f}
  f(r, s) = 2 \cdot (3 r / s + 4 \log s + 7)
\end{align}
as described in~\cite[Theorem~2.4]{ragavan24}.
Note that our expression for the cost differs from the expression in~\cite{ragavan24} in that it depends directly on $K^{(r)}$ instead of estimating $K^{(r)}$ as $\log D/\log{\beta}$ for $\beta = (r + \sqrt{r^2 + 4}) / 2$.

Note furthermore that we have doubled the cost compared to the expression in~\cite{ragavan24}, as the circuit described by Theorem~2.4 must be run twice --- once to compute the product, and once in reverse to clean up intermediary values.

\subsection{Ekerå--Gärtner's extensions}
Ekerå and Gärtner~\cite{eg23-dlog} have extended Regev's algorithm to algorithms for computing discrete logarithms and orders, and for factoring completely via order finding.
In what follows, we refer to these algorithms as Ekerå--Gärtner's extensions of Regev's algorithm~(EGR).

The quantum parts of these extensions are very similar to the quantum part of Regev's original algorithm:
They differ only in that not all of the elements included in the product in the last register of the state in~\refeq{regev-quantum-state} are small.
Instead, in addition to the first~$d$ small elements $a_1, \ldots, a_d$, there are also~$k$ elements $x_1, \ldots, x_k$ that may be arbitrarily selected, for~$k$ a small constant.

The part of the product that depends on the small elements $a_1, \ldots, a_d$ can be computed in exactly the same way as for Regev's original algorithm~\cite{regev23}, with or without the space-saving optimizations of Ragavan and Vaikuntanathan~\cite{rv23v3, ragavan24}.

Hence, with $O(n^{3/2})$~qubits of memory, this part can be computed by calling the quantum circuit in~\refeq{quantum-multiplication} no more than $4 \log D$ times.\footnote{Ragavan~\cite{ragavan24} discusses an option for reducing the cost from $4 \log D$ to $(2 + \epsilon) \log D$.}
Alternatively, it can be computed by calling the multiplication circuit no more than ${f(r, s) \cdot K^{(r)}}$ times while using only~$O(n)$ qubits of memory.

The part of the product that depends on the remaining~$k$ elements $x_1, \ldots, x_k$ can be computed using standard arithmetic in the same way as for the existing variations of Shor's algorithms, see Sect.~\ref{sect:implementing-ekera-hastad}.
Hence, they can be computed by calling the multiplication circuit no more than $2 k \log D$ times.
Furthermore, windowing may be applied to compute this part of the product, again see Sect.~\ref{sect:implementing-ekera-hastad}.
With a window size of~$\cexp$ it is then sufficient to call the multiplication circuit $2 \ceil{(k \log D) / \cexp}$ times.

The full product can thus be computed with $O(n^{3/2})$~qubits of memory by calling the multiplication circuit $4 \log D + 2 \ceil{(k \log D) / \cexp}$ times, or with $O(n)$~qubits of memory by calling the multiplication circuit $f(r, s) \cdot K^{(r)} + 2 \ceil{(k \log D) / \cexp}$ times.

EGR also differ from Regev's original algorithm in how the small elements $a_1, \ldots, a_d$ are selected:
Whereas Regev selects these to be the squares of~$d$ small primes, in EGR they are instead selected to be equal to~$d$ small primes.
This somewhat decreases the cost of computing the full product in the last register in~\refeq{regev-quantum-state}, as the numbers involved in the product are half the size.
Furthermore, this enables somewhat larger values of~$d$ to be used while still ensuring that the cost of computing the partial products~$c_j$ is negligible compared to other costs.

Finally, it should be stated that Ekerå and Gärtner~\cite{eg23-dlog} make a slightly stronger heuristic assumption in their analysis compared to the that made by Regev~\cite{regev23} in his original analysis.
This being said, and as explained in detail by Ekerå and Gärtner~\cite{eg23-dlog}, there is no reason to doubt the validity of either of these heuristic assumptions.
On the contrary, they can be seen to hold for special-form integers and small integers in simulations~\cite{eg23-dlog, eg23-simulations}.
Furthermore, Pilatte~\cite{pilatte24} has proved that a variant of these assumptions hold.

\subsection{The constant~$C$}
\label{sect:C-bound}
As previously explained in Sect.~\ref{sect:methodology}, to compare the cost of EGR to EHS and ES, respectively, we count the number of calls to the multiplication circuit~\refeq{quantum-multiplication} that are required to solve a given problem instance, per run and overall.

For EGR, when using Regev's original arithmetic, the number of large multiplications modulo~$N$ that need to be performed per run depends on the value of $\log D \sim C\sqrt{n}$, and hence on the constant~$C$.

When using Ragavan's generalization of Ragavan and Vaikuntanathan's space-saving optimizations, it also depends on~$r$ and~$s$ via $f(r, s)$ and $K^{(r)} \approx \log D / \log \beta$ where $\beta = (r+\sqrt {r^2+4})/2$.
In Ragavan and Vaikuntanathan's work, $r = 1$, in which case $\beta$ is the golden ratio $\phi = (1+\sqrt 5)/2$.

In~\cite{regev23}, Regev does not explore how to select the constant~$C$ as a function of the dimension~$d$, number of runs~$m$ and quality of the lattice basis reduction.
To explore how to select~$C$, we must examine the heuristic assumption upon which the algorithm relies.

The heuristic assumption varies somewhat between Regev's original algorithm and Ekerå and Gärtner's extensions.
In both cases, the assumption relates to a lattice~$\mathcal L$ that is determined by the small integers $a_1, \ldots, a_d$ and the modulus~$N$.
Ekerå and Gärtner assume for their extensions that this lattice~$\mathcal L$ has a basis that consists of vectors of length at most $T = \exp(O(n/d))$.
This is also a sufficient assumption for Regev's original algorithm.

To analyze which value of~$C$ is sufficient, we consider $T = \exp(\kappa n/d)$ for some constant~$\kappa$.
In~\cite{rv23v3}, for Regev's variant of the heuristic assumption, with the specific choice of $d = \ceil{\sqrt{n}\,}$, it is natural to assume that selecting~$\kappa$ slightly larger than one is sufficient.
It is also natural to assume that the stronger assumption used in the extensions of Ekerå and Gärtner holds for~$\kappa$ slightly larger than one.

In the primary analysis of his algorithm, Regev's used $d = \ceil{\sqrt{n}\,}$ and $m = d + 4$.
However, he also notes that selecting a larger~$d$ and using a computationally more expensive lattice reduction algorithm for the post-processing allows for a more efficient quantum algorithm.
It can furthermore be seen that, even for a fixed value of~$d$, increasing the number of runs~$m$ allows the algorithm to succeed with a smaller value of the constant~$C$.
As such, the choice of~$d$ and~$m$ has a large impact on the efficiency of the algorithm.
Selecting $m \approx d = \ceil{\sqrt{n}\,}$, as previously considered in~\cite{regev23, rv23v3}, is therefore probably not the best choice in practice.

For the classical post-processing to be successful in recovering the solution, Regev's analysis in~\cite[Sect.~5]{regev23} requires that
\[
    (2d+4)^{1/2} \cdot 2^{d+2} \cdot (d+5)^{1/2} \cdot T < \dfrac{\sqrt{2}R}{6d} \cdot (\det \mathcal L)^{-1/m}
\]
for the special case of $m = d + 4$.
It can be seen that for arbitrary~$m$ this requirement corresponds to
\begin{align}\label{eq:post-process-requirement}
    (m+d)^{1/2} \cdot 2^{(m+d)/2} \cdot (m+1)^{1/2} \cdot T < \dfrac{\sqrt{2}R}{6d} \cdot (\det \mathcal L)^{-1/m}
\end{align}

Inserting $T = 2^{\kappa n/d}$, $R = 2^{C\sqrt{n}}$ and $\det{\mathcal L} < 2^n$ into~\refeq{post-process-requirement} leads to the requirement
\[
    (m+d)^{1/2} \cdot 2^{\kappa n/d + (m+d)/2} \cdot (m+1)^{1/2} < \dfrac{\sqrt{2}}{6d} 2^{C\sqrt n - n/m}
\]
which, by taking logarithms on both sides, leads to
\[
    C > \dfrac{\kappa \sqrt{n}}{d} + \dfrac{m+d}{2\sqrt{n}} + \dfrac{\sqrt{n}}{m} + o(1)
\]
where, as previously mentioned, the heuristic assumption should reasonably hold for $\kappa \approx 1$.
Letting $\kappa = 1 + o(1)$, this lower bound on~$C$ is minimized when $m \approx d \approx \sqrt{2n}$, which leads to the conclusion that $C > 2\sqrt{2} + o(1)$ is required.

However, we believe that in practice it is possible to select a value of~$C$ that is significantly smaller than $2\sqrt{2}$.
This is partly due to the fact that LLL~\cite{lll} performs significantly better in practice than Regev's analysis predicts, since it relies on the provable properties of LLL.

Furthermore, the post-processing can also use a computationally more expensive lattice reduction algorithm, such as BKZ~\cite{bkz}, allowing an even smaller value of~$C$ to be used.
This in turn reduces the cost of the quantum algorithm.

\subsubsection{Using better lattice reduction}
The factor~$2^{(m+d)/2}$ in~\refeq{post-process-requirement} is due to Regev's analysis relying on the provable properties of LLL.
In particular, he relies on the Gram--Schmidt norm of two successive vectors of an LLL-reduced basis decreasing by at most a factor $\sqrt 2$, giving rise to the factor $2^{(m+d)/2}$ for the full $(m+d)$-dimensional lattice.
Heuristic analyses of the performance of lattice reduction algorithms often rely on the so-called Geometric Series Assumption (GSA), under which the Gram--Schmidt norm of successive vectors decrease by some factor~$\gamma$.
Under the GSA, we can thus replace the factor $2^{(m+d)/2}$ in Regev's algorithm by a factor~$\gamma^{m+d}$, with~$\gamma$ based on the heuristically estimated performance of the lattice reduction algorithm used.

The performance of lattice reduction algorithms is not typically expressed directly in terms of the factor~$\gamma$ that is relevant for our analysis.
Instead, the root-Hermite factor~$\delta$ is often used, with the shortest vector~$\vect b$ in a reduced basis of an $n$-dimensional lattice~$\mathcal L$ having norm at most $\delta^{n} \cdot (\det{\mathcal L})^{1/n}$.
As the determinant of a lattice is equal to the product of the Gram--Schmidt norms of the vectors in any basis for the lattice, under the GSA we have that
\[
    \det \mathcal L
    =
    \norm{\vect b}^n \prod_{i \, = \, 0}^{n-1} \gamma^{-i}
    =
    \norm{\vect b}^n \gamma^{-(n^2-n)/2},
\]
and thus
\[
    \norm {\vect b}
    =
    \gamma^{(n-1)/2} \cdot (\det \mathcal L)^{1/n}
    \approx
    \gamma^{n/2} \cdot (\det \mathcal L)^{1/n},
\]
and we therefore heuristically estimate $\gamma$ as $\delta^2$.
To estimate~$\delta$ for BKZ~\cite{bkz} with varying block sizes, we use the formula of Chen~\cite{chen2013} with special handling for small block sizes, in the same way as is done in the lattice estimator~\cite{concrete-hardness, lattice-estimator}.

Using~$\gamma$ as determined by the heuristically estimated performance of the lattice reduction employed, the bound on~$C$ then becomes
\begin{align}\label{eq:C-bound}
    C > \dfrac{\sqrt{n}}{d} + \log{\gamma}\left(\dfrac{m+d}{\sqrt{n}}\right) + \dfrac{\sqrt{n}}{m} + o(1)
\end{align}
with $\kappa = 1 + o(1)$.
From this, we can see that the lower bound on~$C$ is minimized when ${m \approx d \approx \sqrt{n/\log \gamma}}$, leading to the requirement that ${C > 4 \sqrt{\log \gamma} + o(1)}$.

\subsection{Selecting parameters}
\label{sect:regev-parameters}
To optimally select the parameters~$d$, $m$, $r$ and $s$ for a given $n$, we need a method for estimating the least value of the constant~$C$ that guarantees a sufficiently high success probability in the classical post-processing.
For this purpose, we use the bound~\refeq{C-bound} in Sect.~\ref{sect:C-bound} whilst ignoring the $o(1)$-term.

A reasonable objection to this approach is that the $o(1)$-term may potentially be significant for the problem instances we consider.
However, we find that the least~$C$ yielded by the bound when ignoring the $o(1)$-term agrees quite well with~$C$ as estimated by our simulator in~\cite{eg23-simulations}.
Based on this observation, we conclude that the $o(1)$-term may be neglected for the problem instances we consider.

By using better lattice reduction, and selecting larger~$d$ and~$m$, the constant~$C$ can be decreased, resulting in a lower cost for the quantum circuit.
If one ignores the cost of the lattice reduction, and consider only the per-run cost, $m$ can be selected arbitrarily large.
This is in contrast to~$d$ which is still restricted.

For our estimates of the cost of the various extensions of Regev's algorithm, we have ignored the cost of computing the $c_j$, but this is only reasonable for as long as $a_1^r\cdot \ldots\cdot a_d^r < N$, see Sect.~\ref{sect:rv-optimizations}, and this restricts~$d$ and~$r$.
Furthermore, the space usage of the quantum circuit increases when~$d$ increases, imposing an additional cost, not captured in our model, of increasing~$d$.

For given~$n$ and~$r$, there is hence a maximal value~$d_{\max}$ for~$d$ such that if~$d$ is larger than~$d_{\max}$, the cost of computing the~$c_j$ is no longer negligible.
Assuming that $d_{\max} < \sqrt{n/\log \gamma}$, we have by simplifying~\refeq{C-bound} that the least~$C$ is given by
\[
    C > \dfrac{\sqrt{n}}{d_{\max}} + \log{\gamma}\dfrac{d_{\max}}{\sqrt{n}} + 2 \sqrt{\log \gamma}
\]
when $m = \sqrt{n/\log \gamma}$.
The least~$C$ is still close to this value if~$m$ is rounded off.
To obtain an optimal parameterization, we hence select
\begin{align*}
    d = \min \left( d_{\max}, \round{\sqrt{n/\log \gamma}} \right)
    \quad \text{ and } \quad
    m = \round{\sqrt{n/\log \gamma}},
\end{align*}
and the least possible~$C$ as given by~\refeq{C-bound} when ignoring the $o(1)$-term.

For given~$n$ and~$r$, we can thus find optimal values for~$d$ and~$m$.
To select~$r$ and~$s$ optimally given $n$, we exhaustively search through reasonable values and select the combination that minimizes the cost in our metric, namely $f(r, s) \cdot K^{(r)}$.

\section{Cost comparisons and results}
\label{sect:cost-comparisons-results}
In App.~\ref{app:tables}, we present several cost comparisons between the existing variations of Shor's algorithms on the one hand, and EGR with the space-saving optimizations of Ragavan and Vaikuntanathan on the other.
In particular, we focus on cryptographically relevant instances of the RSA integer factoring problem~(IFP) and the discrete logarithm problem~(DLP).

The most interesting comparison is between Regev's algorithm, with all currently available improvements and optimizations applied, and somewhat optimized versions of the existing variations of Shor's algorithms.
Some details differ between the IFP and DLP, and between the different problem instances for the DLP that we consider, whilst other overarching aspects remain the same.

In particular, Ragavan's generalization~\cite{ragavan24} of the aforementioned space-saving optimizations yields the lowest cost in our metric.
Furthermore, to optimize the cost of Regev's algorithm in our metric, the best lattice reduction algorithm that it is practically feasible to use should be employed for the post-processing.
Given an estimate of the performance of this lattice reduction algorithm, the optimal cost in our metric is achieved by selecting~$d$, $m$, $r$ and~$s$ as described in Sect.~\ref{sect:regev-parameters}.

In practice, the best lattice reduction is achieved by the BKZ algorithm~\cite{bkz} with as large a block size as it is practically feasible to select.
It is not clear exactly what the limit is for the block size.
So as to avoid overestimating the cost of Regev's algorithm, given that we intentionally seek to bias our comparisons to be in its favor, we select a block size of 200 in our primary comparisons.
This is clearly larger than what is practically feasible today, yet not too large of an exaggeration.
We heuristically estimate the performance of BKZ, and select the minimum~$C$ with precision~$0.01$ that allows the post-processing to succeed, again see Sect.~\ref{sect:regev-parameters}.

In addition to our primary comparisons, to illustrate the benefits of the different optimizations that we apply to Regev's algorithm, and of windowing, we provide a greater range of comparisons for RSA.

\subsection{Comparisons for RSA}
In this section, we compare Regev's algorithm that solves the IFP to EHS that solves the RSA IFP.
In particular, we focus on Ekerå--Gärtner's extension of Regev's algorithm to factoring via order finding~\cite[App.~A]{eg23-dlog} since it allows for slightly better flexibility in how parameters are selected.
In turn, this allows the algorithm to achieve a slightly better efficiency in our cost metric.

We consider $n$-bit RSA integers~$N$ for $n \in \{2048, 3072, 4096, 6144, 8192\}$ so as to model RSA-2048 up to and including RSA-8192.
By RSA integer we mean an integer with two random prime factors of identical bit lengths.

For EHS, we consider both the case where we do not make tradeoffs and solve in a single run, and the case where we make tradeoffs with a reasonably large tradeoff factor.
For further details, see App.~\ref{app:tables}.

To start off, we first provide a baseline comparison in Sect.~\ref{sect:rsa-ifp-lll} between Regev's algorithm parameterized as originally proposed by Regev, and EHS without windowing.
For all other comparisons in Sects.~\ref{sect:rsa-ifp-lll-scaled-up-d-m}--\ref{sect:rsa-ifp-perfect-reduction}, we apply windowing with a window size of $\cexp = 10$ to EHS.

The comparisons in Sects.~\ref{sect:rsa-ifp-lll-scaled-up-d-m}--\ref{sect:rsa-ifp-perfect-reduction} differ only with respect to how Regev's algorithm and its extensions are parameterized.
We select optimal~$d$ and~$m$ as described in Sect.~\ref{sect:regev-parameters}, and explore different choices of lattice reduction algorithms, and of using the generalized Fibonacci-based exponentiation proposed by Ragavan.

\subsubsection{A basic baseline comparison}
\label{sect:rsa-ifp-lll}
In App.~\ref{app:rsa-ifp-lll}, for Regev's algorithm, we use LLL, and let $d = \ceil{\sqrt{n}\,}$ and ${m = d + 4}$, as originally proposed by Regev.
Furthermore, we use Ragavan's generalization of the space-saving optimizations with ${r = 1}$ leading to~$20 K$ operations in the form of large multiplications modulo~$N$ being required per run.

As may be seen in Tab.~\ref{table:rsa-ifp-lll} in App.~\ref{app:rsa-ifp-lll}, Regev's algorithm performs approximately between a factor two to four fewer operations per run than EHS when not making tradeoffs.
However, EHS requires significantly fewer operations overall when not making tradeoffs, as it then only requires a single run.
Hence, if the goal is to optimize the overall work, not making tradeoffs is typically preferable.

By making tradeoffs, EHS has an advantage over Regev's algorithm, both per run and overall, for RSA-2048.
For RSA-3072, the per-run cost is essentially on par, but EHS has the overall advantage.
For RSA-4096 up to and including RSA-8192, EHS retains the overall advantage, but it has a per-run disadvantage.

Asymptotically, Regev's algorithm has the per-run advantage, and this is consistent with the behavior we observe in Tab.~\ref{table:rsa-ifp-lll}.
This is the case for all the parameterization of Regev's algorithm that we consider for the RSA IFP, but the break-even point differs between the parameterizations.

The per-run cost for Regev's algorithm is essentially determined by the value of the constant~$C$, see Sect.~\ref{sect:C-bound}.
As may be seen in Tab.~\ref{table:rsa-ifp-lll}, and as previously pointed out in~\cite{eg23-dlog}, it suffices to select $C \approx 2$ when the algorithm is parameterized as in this baseline comparison.
By selecting a better parameterization, and potentially also using a better lattice reduction algorithm, the value of~$C$ can be decreased.

\subsubsection{Using LLL and $r = 1$}
\label{sect:rsa-ifp-lll-scaled-up-d-m}
In App.~\ref{app:rsa-ifp-lll-scaled-up-d-m}, we use LLL, select $r = 1$, and select optimal~$d$ and~$m$, for EGR.
Furthermore, we apply windowing with ${\cexp = 10}$ for EHS from this point on.

As may be seen in Tab.~\ref{table:rsa-ifp-lll-scaled-up-d-m} in App.~\ref{app:rsa-ifp-lll-scaled-up-d-m}, selecting $C \approx 1$ for EGR is sufficient with this parameterization.
The cost per run for EGR with this parameterization is thus approximately halved compared to the baseline.
Furthermore, due to selecting a larger~$m$, the overall cost for EGR increases by approximately a factor of two compared to the baseline, leading to EHS having a significant overall advantage --- but our goal in this comparison is to optimize for the per-run cost.

This being said, even though the cost for EGR decreases by a factor of two compared to the baseline, the per-run gap to EHS in fact increases, thanks to windowing having been applied.
Windowing reduces both the per-run cost and the overall cost of EHS by approximately a factor of $\cexp = 10$.
EHS therefore now has the per-run advantage for RSA-2048 up to and including RSA-8192.
This underscores how important an optimization windowing is in practice.

From this point on, we apply no further optimizations to EHS, so the costs in Tab.~\ref{table:rsa-ifp-lll-scaled-up-d-m} for EHS are the costs that EGR must go below to achieve an advantage.

\subsubsection{Using BKZ-200 and $r = 1$}
\label{sect:rsa-ifp-bkz-200-scaled-up-d-m}
In App.~\ref{app:rsa-ifp-bkz-200-scaled-up-d-m}, we replace LLL with BKZ-200, select $r = 1$, and optimal~$d$ and~$m$.

As may be seen in Tab.~\ref{table:rsa-ifp-bkz-200-scaled-up-d-m} in App.~\ref{app:rsa-ifp-bkz-200-scaled-up-d-m}, with this parameterization, it is sufficient to select~$C \approx 0.5$.
Thus, if BKZ-200 can be used for post-processing instead of LLL, the per-run cost of the quantum algorithm can be decreased by approximately a factor of two.
However, this reduction comes at the cost of approximately doubling the number of runs required, and the overall cost of the quantum algorithm thus remains approximately the same.

Note that, for most of the problem instances, the possible choices of~$d$ are limited as explained in the analysis in Sect.~\ref{sect:regev-parameters}.
This limitation becomes less and less restrictive as the problem size increases.
In Tab.~\ref{table:rsa-ifp-bkz-200-scaled-up-d-m}, it is only for RSA-8192 that this limitation does not prevent us from selecting~$d$ optimally.
It is because of this limitation that there is some variance in which values of~$C$ can be selected for the different problem instances.

By using BKZ-200 in the post-processing, EGR now again outperforms EHS per run when not making tradeoffs, for RSA-4096 up to RSA-8192.
For RSA-4096, the two algorithms are essentially on par.
Note however that, for all problem instances, including the aforementioned instances, the overall cost of EGR is more than $400$~times greater than that of EHS when not making tradeoffs.

By making tradeoffs, EHS outperforms EGR, both per run and overall, for RSA-2048 up to and including RSA-8192.
For EGR to achieve a per-run advantage compared to EHS when making tradeoffs, even for the largest problem instances we consider, its cost must decrease by almost a factor of two.
The first multiple of 1024~bits for which EGR achieves a per-run advantage with this parameterization is for RSA-27648, corresponding to a problem instance with a classical strength level of 336~bits in the NIST model~\cite[Sect.~7.5 on p.~126]{fips-140-2-guidance}.

\subsubsection{Using LLL and optimal~$r$}
\label{sect:rsa-ifp-lll-optimal-r}
In App.~\ref{app:rsa-ifp-lll-optimal-r}, we once again use LLL, select optimal~$r$ and~$s$, and optimal~$d$ and~$m$.

For all the problem instances we consider, it is optimal to select $r = 4$, since the greatest cost reduction is achieved when~$r$ is a power of two, and since selecting $r = 8$ imposes too strict a limit on~$d$ for these instances.

Note furthermore that, for all of the problem instances we consider, the choice of~$d$ is limited as explained in the analysis in Sect.~\ref{sect:regev-parameters} due to us having selected optimal $r > 1$.
It is because of this limitation that there is quite a large variance in which values of~$C$ can be selected for the different problem instances.

As may be seen in Tab.~\ref{table:rsa-ifp-lll-optimal-r} in App.~\ref{app:rsa-ifp-lll-optimal-r}, the effect of picking optimal~$r$ and~$s$ in Ragavan's generalization of the space-saving optimizations has a nearly equivalent effect on the cost per run to that of using BKZ-200 instead of LLL, see Tab.~\ref{table:rsa-ifp-bkz-200-scaled-up-d-m} in App.~\ref{app:rsa-ifp-bkz-200-scaled-up-d-m} and compare.
Meanwhile, the overall cost is decreased by almost a factor of two compared the cost when selecting $r = 1$ and optimal~$d$ and~$m$.

EGR again outperforms EHS per run when not making tradeoffs, for RSA-6144 and RSA-8192, but its overall cost is at least a factor $240$~times greater.
By making tradeoffs, EHS outperforms EGR, both per run and overall, for RSA-2048 up to and including RSA-8192.

Note than an important distinction between the parameterization in this section and the one in and Sect.~\ref{sect:rsa-ifp-bkz-200-scaled-up-d-m} is that the post-processing in this section is feasible to execute in practice, whereas the post-processing in Sect.~\ref{sect:rsa-ifp-bkz-200-scaled-up-d-m} is impractical.

\subsubsection{Using BKZ-200 and optimal~$r$}
\label{sect:rsa-ifp-bkz-200-optimal-r}
In App.~\ref{app:rsa-ifp-bkz-200-optimal-r}, we use BKZ-200, select optimal~$r$ and~$s$, and optimal~$d$ and~$m$.

For all the problem instances we consider, it is optimal to select $r \in \{2, 4\}$ when using BKZ-200 for the post-processing.
As in Sect.~\ref{sect:rsa-ifp-lll-optimal-r}, using a larger~$r$ imposes too strict a limit on the choice of~$d$ for it to be beneficial.
Furthermore, with better lattice reduction, increasing~$d$ has a larger impact on how small~$C$ can be selected.
This explains why, compared to when using LLL for the post-processing, a smaller~$r$ and larger~$d$ is sometimes selected when using BKZ-200.

As may be seen in Tab.~\ref{table:rsa-ifp-bkz-200-optimal-r} in App.~\ref{app:rsa-ifp-bkz-200-optimal-r}, by using all available optimizations, and BKZ-200 for the post-processing, we achieve a significantly lower per-run cost than in any of the previously considered parameterizations.
Compared to using LLL with optimal~$r$, the per-run cost decreases by slightly less than a factor of~$4/3$.
The improvement achieved by using better lattice reduction is thus smaller in this case where we use optimal~$r$ compared to the case when $r = 1$, see Sect.~\ref{sect:rsa-ifp-bkz-200-scaled-up-d-m}.

As previously mentioned, the use of larger~$r$ imposes a stricter limit on the value of~$d$.
This limit on~$d$ is the reason for why the value of the constant~$C$ varies quite heavily between the different problem instances we consider.
It is also because of this restriction on~$d$ that the efficiency gain of using better lattice reduction is smaller compared to the case when $r = 1$.

This parameterization of EGR achieves a per-run advantage against EHS, when not making tradeoffs, for RSA-3072 and larger problem instances.
While it never achieves an advantage against EHS when making tradeoffs, for RSA-6144 and RSA-8192, the per-run advantage of EHS is below a factor of two.

The first multiple of 1024~bits for which EGR achieves a per-run advantage with this parameterization is for RSA-13312, corresponding to a problem instance with a classical strength level of 248~bits in the NIST model~\cite[Sect.~7.5 on p.~126]{fips-140-2-guidance}.

\subsubsection{Using perfect reduction and optimal~$r$}
\label{sect:rsa-ifp-perfect-reduction}
In App.~\ref{app:rsa-ifp-perfect-reduction}, we use perfect lattice reduction, as modelled by letting $\gamma \rightarrow 1$, select optimal~$r$ and~$s$, and optimal~$d$ and~$m$.
Note that this implies that $m \rightarrow \infty$, leaving the bound on~$C$ as~$C > \sqrt{n} / d$.
Note furthermore that when using this model of perfect lattice reduction, it is no longer beneficial to select $r > 1$, as doing so leads to a stronger restriction on the value of~$d$.
With perfect lattice reduction, the benefit of being able to select a smaller~$C$ as made possible by increasing~$d$ outweighs the benefit of increasing~$r$.

Needless to say, this not a realistic cost comparison since perfect lattice reduction is completely impractical.
Rather, its use essentially corresponds to the situation in the curious corollary by Regev in the introduction of~\cite{regev23}, that states that a quantum circuit for factoring with essentially linear size in~$n$ exists if lattice problems are easy to solve classically.
In our model, this is captured by~$C$ not being constant.
Rather, increasingly smaller values of~$C$ can be used as the size of the problem instance grows larger.

For this cost comparison, we consider two different limits on how~$d$ is selected in order to illustrate the advantage of factoring via order finding with EGR compared to factoring via Regev's original algorithm.

In EGR, the small $a_1, \ldots, a_d$ are the first~$d$ primes, while in Regev's original algorithm, they are the squares of the first~$d$ primes.
This results in Regev's algorithm imposing a stricter limit on the choice of~$d$, which, depending on which lattice reduction is used for the post-processing, may have a significant effect on the performance.
The better the lattice reduction, the larger the impact, and there is therefore a significant performance difference between EGR and Regev's original algorithm.
This is evident from Tab.~\ref{table:rsa-ifp-perfect-reduction} in App.~\ref{app:rsa-ifp-perfect-reduction} where perfect lattice reduction is used, and~$d$ is picked optimally to respect the limits imposed by Regev's algorithm and EGR, respectively, when comparing to EHS with tradeoffs.

As may be seen in Tab.~\ref{table:rsa-ifp-perfect-reduction} in App.~\ref{app:rsa-ifp-perfect-reduction}, even in this extremely biased comparison EHS retains a per-run advantage against Regev's original algorithm up to RSA-6144 where the algorithms are essentially on par.
When instead comparing against EGR, EHS is essentially on par for RSA-4096.

Note that we do not provide a comparison of the overall costs of the algorithms in Tab.~\ref{table:rsa-ifp-perfect-reduction}.
This is because the overall number of operations tends to infinity with~$m$ for Regev's algorithm and EGR, giving EHS an infinite overall advantage.

\subsection{Comparisons for discrete logarithms}
In addition to the above comparison for the RSA IFP, we provide analogous comparisons for the DLP in $r$-order subgroups to~$\mathbb Z_N^*$, for~$N = 2ur + 1$ a large $n$-bit prime, and~$r$ a prime.
We use the model of NIST, as given in~\cite[Tab.~25--26 in App.~D on p.~133]{nist-sp-800-56a} and~\cite[Sect.~7.5 on p.~126]{fips-140-2-guidance}, to estimate the classical strength level~$z$ of~$N$.
As for RSA, we consider $n \in \{ 2048, 3072, 4096, 6144, 8192 \}$.
We consider both safe-prime groups for which $u = 1$, and Schnorr groups for which~$r$ is of length $2z$~bits.
Both of these groups have a classical strength level of $z$~bits.

In the case of safe-prime groups, we consider the short DLP where the logarithm is of length $2z$~bits, and the general DLP where the logarithm is on $[0, r)$.
Again, both of these problems have a classical strength level of $z$~bits.
Note that cryptographic applications are typically based on the short DLP since it is much more efficient, but we nevertheless also include the general DLP in our comparison.

When using EGR to compute discrete logarithms, the algorithm cannot directly leverage that the $r$-order subgroup is small, as is the case for Schnorr groups, nor that the logarithm is short, so its cost is the same for all three parameterizations.
We compare EGR to ES for the general DLP in Schnorr groups and safe-prime groups, and to EHS for the short DLP in safe-prime groups.

Note that in~\cite{eg23-dlog}, several extensions of Regev's algorithm are given for the DLP that use~$k \in \{1, 2\}$, for~$k$ as defined in Sect.~\ref{sect:regev-extensions}.
In our comparison, for EGR, we use an extension with $k = 1$, necessitating either doubling the number of runs or performing some pre-computation for the specific choice of generator.
We do not include these costs in the cost we report for EGR.
The high-level outcome of the comparison would not be affected by choosing an extension with $k = 2$, but we select one with $k = 1$ to minimize the cost in our metric.

\subsubsection{General DLP in safe-prime groups}
\label{sect:dlp-safe-prime-group-general}
As may be seen in Tab.~\ref{table:dlp-safe-prime-group-general} in App.~\ref{app:dlp-safe-prime-group-general}, EGR achieves a slight per-run advantage for 6144-bit and 8192-bit moduli, when comparing against ES and making tradeoffs.
Despite all available optimization having been applied to EGR, for 2048-bit moduli, ES almost has the advantage, even when solving in a single run without making tradeoffs.
As for the overall cost, ES has the advantage by more than factor~$240$ when not making tradeoffs, for all of the problem instances considered.

Note that the reason for why EGR achieves an advantage for the general DLP, but not for the RSA IFP, is not because of differences between these extensions.
Rather, it is because EHS for the RSA IFP achieves a reduction in the total control register length by using an efficient reduction from the RSA IFP to the short DLP.

Note furthermore that, as stated above, the cryptographically relevant cases are short discrete logarithms in safe-prime groups, and discrete logarithms in Schnorr groups, respectively, see the next two sections.

\subsubsection{Short DLP in safe-prime groups}
\label{sect:dlp-safe-prime-group-short}
As may be seen in Tab.~\ref{table:dlp-safe-prime-group-short} in App.~\ref{app:dlp-safe-prime-group-short}, when the logarithm is short, EHS has a considerable advantage, both per run and overall, for 2048-bit moduli up to and including 8192-bit moduli.

This is because EHS is specifically tailored to leverage the fact that the logarithm is short, whereas the complexity of EGR depends on the size of~$\mathbb Z_N^*$.
It is hence the same irrespective of whether the logarithm is short or full length.
The advantage becomes even greater due to the use of windowing, but even without windowing EHS would have an advantage when making tradeoffs.

\subsubsection{DLP in Schnorr groups}
\label{sect:dlp-schnorr-group}
As may be seen in Tab.~\ref{table:dlp-schnorr-group} in App.~\ref{app:dlp-schnorr-group}, for Schnorr groups, ES has a considerable advantage, both per run and overall, for 2048-bit moduli up to an including 8192-bit moduli.

Again, this is because ES is specifically tailored to leverage the fact that the subgroup is small, whereas the complexity of EGR depends on the size of~$\mathbb Z_N^*$.
It is hence the same irrespective of whether~$g$ generates~$\mathbb Z_N^*$, or a small subgroup of~$\mathbb Z_N^*$.
Again, the advantage becomes even greater due to the use of windowing, but even without windowing ES would have an advantage when making tradeoffs.

\subsubsection{On the asymptotic advantage of Ekerå--Gärtner}
In analogy to the situation for the RSA IFP, EGR achieves an asymptotic advantage over ES for the general~DLP.
This may be seen in Tab.~\ref{table:dlp-safe-prime-group-general} in App.~\ref{app:dlp-safe-prime-group-general} where the advantage of ES decreases with the size of the problem instance.

For the short DLP, and the DLP in Schnorr groups, the situation is quite different, however.
In these cases, the total exponent length in EHS and ES is a small multiple of the strength level $z = O(n^{1/3} \log^{2/3} n)$, see the NIST model~\cite[Sect.~7.5 on p.~126]{fips-140-2-guidance}.
The number of large multiplication operations modulo~$N$ that need to be performed is twice the total exponent length.
Meanwhile, $\Theta(n^{1/2})$ such operations need to be performed in EGR.

Note that in Tab.~\ref{table:dlp-safe-prime-group-short}--\ref{table:dlp-schnorr-group} in App.~\ref{app:dlp-safe-prime-group-short}--\ref{app:dlp-schnorr-group}, the advantage of EHS and ES does decrease with the size of the problem instance, but this phenomenon is not indicative of the asymptotic behavior.

\section{Summary and conclusion}
\label{sect:summary-conclusion}

Regev's factoring algorithm~\cite{regev23} with the space-saving optimizations of Ragavan and Vaikuntanathan~\cite{rv23v3}, as generalized by Ragavan~\cite{ragavan24}, does not achieve an advantage over the existing state-of-the-art variations of Shor's algorithms for crypto\-graphically relevant problem instances.
Further optimizations are required for it to achieve such an advantage.
The same holds true for Ekerå--Gärtner's extensions to computing discrete logarithms, and to factoring via order finding.

This when using the number of large multiplications modulo~$N$ per run as the cost metric in a black-box model, and ignoring the fact that the space usage, and the number of runs required, is larger for Regev's algorithm and Ekerå--Gärtner's extensions.
If we would instead have used a different cost metric that also accounts for the space usage, such as the spacetime volume, the difference in the advantage would have been even more pronounced.

It would also have been more pronounced if we had attempted to account for the overhead of quantum error correction.
The fact that Regev's algorithm and Ekerå--Gärtner's extensions typically require many more runs than the existing state-of-the-art algorithms drives up the cost of the error correction.
This is because the error correction has to be parameterized so as to ensure that all runs are correct with sufficiently high probability.
Alternatively, the erroneous runs can be filtered out before the post-processing, or the erroneous runs jointly post-processed alongside the correct runs, as proposed by Ragavan and Vaikuntanathan~\cite{rv23v3}, and by Ekerå and Gärtner~\cite{eg23-dlog}, respectively.
However, this requires increasing~$C$, driving up the per-run cost of the algorithm in our metric, and it requires making even more runs.

Treating the multiplication circuit as a black box precludes the use of known optimizations that would benefit the existing state-of-the-art variations of Shor's algorithm, but that are not directly applicable to Regev's algorithm and Ekerå--Gärtner's extensions, further biasing the comparison in our cost metric.
In practice, choosing the multiplication circuit is non-trivial.
The same circuit need not necessarily be optimal for all of the aforementioned algorithms, depending on what other optimizations can be applied, but going into details on the implementation of the multiplication circuit is beyond the scope of this work.

In summary, the cost metric we use in this work is deliberately biased in favor of Regev's algorithm and Ekerå--Gärtner's extensions.
This is not an issue, however, when our comparison shows that the existing state-of-the-art variations of Shor's algorithm have the advantage.
We can then still draw valid conclusions.

Regev's algorithm~\cite{regev23} without the space-saving optimizations does achieve an advantage over Ekerå--Håstad's algorithm for cryptographically relevant instances of the RSA IFP.\footnote{The per-run cost of Regev's original algorithm and Ekerå--Gärtner's extensions without the space-saving optimizations is $4 \log D$ in our metric, for $D$ as in the tables in App.~\ref{app:tables}. As previously noted, Ragavan~\cite{ragavan24} discusses an option for reducing the cost from $4 \log D$ to $(2 + \epsilon) \log D$.}
This when using the aforementioned biased cost metric, and when considering a standard implementation of Ekerå--Håstad's algorithm with windowing but no other significant optimizations.
The same holds true for Ekerå--Gärtner's extensions of Regev's algorithm, for the general DLP in safe-prime groups --- but not for the short DLP in such groups, or for the DLP in Schnorr groups, which are arguably the two cryptographically relevant problems to consider.

Hence, if we envisage non-computational quantum memory\footnote{Memory to which we can swap out a state for an extended period of time and then swap it back in at a later time when it is needed for the uncomputation in Regev's arithmetic.} to be cheap, then Regev's factoring algorithm may have an advantage over the existing state-of-the-art algorithms for cryptographically relevant instances of the RSA IFP.
Note however that even in our biased cost metric, where we ignore the space usage, there is only a per-run advantage --- not an overall advantage.

A key reason for why the existing state-of-the-art variations of Shor's algorithm outperform Regev's algorithm in our comparisons is that many of the techniques used to derive and optimize these variations do not translate directly to Regev's algorithm.
Regev's algorithm is still quite new, however, and more optimizations may become available as time progresses.
A more detailed cost comparison may then be warranted to establish which algorithm has the advantage, but for now, the above biased comparison suffices, when not considering space to be cheap.

\section*{Acknowledgements}
We are grateful to Johan Håstad for his comments on early versions of this manu\-script, and to Oded Regev's for his generous support.
We thank Craig Gidney, Seyoon Ragavan and Vinod Vaikuntanathan for useful discussions.

This work was supported by the KTH Centre for Cyber Defence and Information Security~(CDIS), and funded and supported by the Swedish NCSA that is a part of the Swedish Armed Forces.
Computations were enabled by resources provided by the National Academic Infrastructure for Supercomputing in Sweden (NAISS) at PDC at KTH partially funded by the Swedish Research Council through grant agreement no.~2022-06725.

\appendix
\section{Tables}
\label{app:tables}
In this appendix, we tabulate cost comparisons between Regev's variation~\cite{regev23} of Shor's algorithms~\cite{shor94, shor97}, and Ekerå--Gärtner's extensions~\cite{eg23-dlog} thereof, on the one hand, and Ekerå--Håstad's~\cite{ekera-hastad} and Ekerå's~\cite{ekera-revisiting} variations of Shor's algorithms~\cite{shor94, shor97} on the other.
In the tables, we inherit notation from the aforementioned works, sometimes leading to the same symbol being used to denote different quantities depending on where in the table it is used.
To prevent misunderstandings, we describe how the tables were constructed below:

\begin{itemize}
  \item {\bfseries For Regev's algorithm and Ekerå--Gärtner's extensions (EGR):}

  As in~\cite{regev23, eg23-dlog}, we let $n = \ceil{\log N}$.
  We let~$d$ denote the dimension, and~$m$ the number of runs.
  We let $R = 2^{C \sqrt{n}}$, for~$C > 0$, and ${D = 2^{\lceil \log(2 \sqrt{d} R) \rceil}}$.

  We use Ragavan's generalization~\cite{ragavan24} of the space-saving optimizations of Ragavan and Vaikuntanathan~\cite{rv23v3}.
  We let~$r$ and~$s$ be as defined by Ragavan~\cite{ragavan24}.

  We let $K^{(r)}$ be maximal such that $G^{(r)}_{K^{(r)}} \le D$ as in~\cite[Sect.~3.2]{ragavan24}.
  If $r > 1$, we tabulate~$K^{(r)}$ and~$r$.
  Otherwise, if $r = 1$, we only tabulate $K = K^{(1)}$.

  We do not tabulate~$s$.
  Instead, we pick~$s$ as function of~$r$:
  If $r = 1$, we must pick $s = 1$.
  Otherwise, if $r > 1$ and a power of two, we pick $s = r/2$.
  Otherwise, we pick~$s$ as the greatest power of two that divides~$r$.
  This is the optimal way to pick~$s$ in our cost metric as it minimizes~$f(r, s)$, see Sect.~\ref{sect:rv-optimizations}.

  The number of large multiplications modulo~$N$ of the form
  \begin{align}
    \label{eq:app-form}
    \ket{u, v, t, 0^S} \rightarrow \ket{u, v, (t + uv) \text{ mod } N, 0^S}
  \end{align}
  that need to be performed per run is $f(r, s) \cdot K^{(r)}$, see Sect.~\ref{sect:rv-optimizations}.
  Overall, the number of such operations that need to be performed is hence $f(r, s) \cdot K^{(r)} \cdot m$.
  We tabulate these quantities in the columns denoted ``\#ops''.
  (In Tab.~\ref{table:rsa-ifp-perfect-reduction}, where $m \rightarrow \infty$, we tabulate only the number of operations per run.)

  \item {\bfseries For Ekerå--Håstad's variation of Shor's algorithm (EHS):}

  As in~\cite{ekera-hastad}, we let~$m$ be an upper bound on the bit length of the short discrete logarithm~$d$.
  For the RSA IFP, we have that $m = \ceil{\log N} / 2 - 1$, whereas $m = 2z$ for the DLP in safe-prime groups, for~$z$ the strength level in bits.

  We consider both the case of solving for~$d$ in a single run via~\cite{ekera-short-success} with $\Delta = 30$ and $\ell = m - \Delta$, and of making tradeoffs with tradeoff factor $s > 1$ and solving via~\cite{ekera-pp} in $n \ge s$ runs with $\ell = \ceil{m / s}$.

  In the former case, the success probability $\ggg 99\%$ , see~\cite[Tab.~1]{ekera-short-success}.
  In the latter case, we pick~$s$ and~$n$ from~\cite[Tab.~3]{ekera-pp}, leading to a success probability $\ge 99\%$ without enumerating.
  Other parameterizations are of course possible.

  We use standard arithmetic, with windowing with a ${\cexp = 10}$~bit window, and without windowing corresponding to having a $\cexp = 1$~bit window.

  The total exponent length in each run is $m + 2\ell$.
  It follows that number of large multiplications modulo~$N$ of the form~\refeq{app-form} that need to be performed per run is $2 \cdot \ceil{(m + 2 \ell) / \cexp}$.
  Overall, the number of such operations that need to be performed is hence $2n \cdot \ceil{(m + 2 \ell) / \cexp}$.
  We tabulate these quantities in the columns denoted ``\#ops''.

  We define the advantage as the quotient between the number of such operations required by Regev's algorithm, or EGR, and EHS, respectively.
  We tabulate the advantage, per run and overall, in the columns denoted ``adv''.
  (In Tab.~\ref{table:rsa-ifp-perfect-reduction}, we tabulate only the advantage per run.)

  \item {\bfseries For Ekerå's variation of Shor's algorithm (ES):}

  As in~\cite{ekera-revisiting}, we let~$m$ be an upper bound on the bit length of the logarithm~$d$.
  For the general DLP in safe-prime groups, we have that $m = \ceil{\log N} - 1$, whereas we instead have that $m = 2z$ for the DLP in Schnorr groups, for~$z$ the strength level in bits.

  We consider both the case of solving for~$d$ in a single run via~\cite[Sect.~6.1]{ekera-revisiting} with $\varsigma = 0$ and $\ell = m$, and of making tradeoffs with tradeoff factor $s > 1$ and solving via~\cite[Sect.~6.2]{ekera-revisiting} in $n \ge s$ runs with $\ell = \ceil{m / s}$.

  In the former case, the success probability $\ggg 99\%$, provided that one performs a limited search in the classical post-processing, see~\cite[Tab.~1 in App.~A.1]{ekera-short-success}.
  In the latter case, we pick~$s$, $\varsigma$ and~$n$ so as to achieve a reasonable tradeoff and a  success probability $\ge 99\%$ without enumerating, and with $\eta_1 = \ldots = \eta_n = 0$.
  This based on simulations performed with the Qunundrum~\cite{ekera-qunundrum} suite of MPI programs on the Dardel HPE Cray EX supercomputer at PDC at KTH.
  Other parameterizations are of course possible.

  We use standard arithmetic, with windowing with a ${\cexp = 10}$~bit window.

  The total exponent length in each run is $m + \varsigma + \ell$.
  It follows that number of large multiplications modulo~$N$ of the form~\refeq{app-form} that need to be performed per run is $2 \cdot \ceil{(m + \varsigma + \ell) / \cexp}$.
  Overall, the number of such operations that need to be performed is hence $2n \cdot \ceil{(m + \varsigma + \ell) / \cexp}$.
  We tabulate these quantities in the columns denoted ``\#ops''.

  We define the advantage as the quotient between the number of such operations required by Regev's algorithm, or EGR, and ES, respectively.
  We tabulate the advantage, per run and overall, in the columns denoted ``adv''.

  It should be noted that the analysis in~\cite{ekera-revisiting} is heuristic.
  We could instead have based our comparison on the algorithm and analysis in~\cite{ekera-general} that does not require the group order to be known.
  This would have led to a cost profile essentially identical to that for EHS~\cite{ekera-hastad, ekera-pp, ekera-short-success}, but for the fact that somewhat smaller tradeoff factors~$s$ would have had to be selected for the Schnorr groups to ensure that the upper bound on the approximation error in the analysis in~\cite{ekera-general} is sufficiently low.
\end{itemize}

\clearpage

\begin{landscape}
  \subsection{RSA IFP}
  \label{app:rsa-ifp}

  \subsubsection{A basic baseline comparison}
  \label{app:rsa-ifp-lll}
  \begin{table}[h!]
    \begin{center}
      {\setlength{\tabcolsep}{0.45em}
      \renewcommand{\arraystretch}{1.25}
      \begin{tabular}{c|ccccc|r|r|cccc|r S[table-format=1.3] | r S[table-format=2.2]}
        \thickhline
        & \multicolumn{7}{c|}{\bfseries IFP via Regev~\cite{regev23} with~\cite{rv23v3, ragavan24}}
        & \multicolumn{8}{c}{\bfseries RSA IFP via Ekerå--Håstad~\cite{ekera-hastad, ekera-pp, ekera-short-success}} \\
        &
        &&&&& \multicolumn{1}{c|}{\bfseries per run} & \multicolumn{1}{c|}{\bfseries overall}
        &&&&& \multicolumn{2}{c|}{\bfseries per run} & \multicolumn{2}{c}{\bfseries overall}
        \\
        $\ceil{\, \log N \,}$ &
        $d$ & $m$ & $C$ & $\log D$ & $K$ & \#ops & \#ops &
        $m$ & $s$ & $\ell$ & $n$ & \#ops & \multicolumn{1}{r|}{adv} & \#ops & \multicolumn{1}{r}{adv} \\
        \thickhline
        2048 &  46 &  50 & 2.03 &  96 & 138 &  2760 &  138000 & 1023 & -- &  993 &  1 &  6018 &    0.46 &   6018 &   22.9  \\
             &     &     &      &     &     &       &         &      & 17 &   61 & 20 &  2290 &    1.20 &  45800 &    3.01 \\
        \hline
        3072 &  56 &  60 & 2.05 & 118 & 170 &  3400 &  204000 & 1535 & -- & 1505 &  1 &  9090 &    0.37 &   9090 &   22.4  \\
             &     &     &      &     &     &       &         &      & 21 &   74 & 24 &  3366 &    1.01 &  80784 &    2.52 \\
        \hline
        4096 &  64 &  68 & 2.08 & 138 & 199 &  3980 &  270640 & 2047 & -- & 2017 &  1 & 12162 &    0.33 &  12162 &   22.2  \\
             &     &     &      &     &     &       &         &      & 24 &   86 & 27 &  4438 &    0.90 & 119826 &    2.25 \\
        \hline
        6144 &  79 &  83 & 2.07 & 167 & 241 &  4820 &  400060 & 3071 & -- & 3041 &  1 & 18306 &    0.26 &  18306 &   21.8  \\
             &     &     &      &     &     &       &         &      & 31 &  100 & 34 &  6542 &    0.74 & 222428 &    1.79 \\
        \hline
        8192 &  91 &  95 & 2.08 & 193 & 278 &  5560 &  528200 & 4095 & -- & 4065 &  1 & 24450 &    0.23 &  24450 &   21.6  \\
             &     &     &      &     &     &       &         &      & 34 &  121 & 37 &  8674 &    0.64 & 320938 &    1.64 \\
        \thickhline
      \end{tabular}}
    \end{center}
    \caption{Comparison between Regev's algorithm~\cite{regev23} (with~\cite{rv23v3, ragavan24}, LLL, $r = 1$, $d = \ceil{\sqrt{n}\,}$, $m = d + 4$) and Ekerå--Håstad's variation~\cite{ekera-hastad, ekera-pp, ekera-short-success} of Shor's algorithm~\cite{shor94, shor97} for the RSA IFP (with $w = 1$).
    For further information on this table and how to interpret it, see Sect.~\ref{sect:rsa-ifp-lll} and App.~\ref{app:tables}.}
    \label{table:rsa-ifp-lll}
  \end{table}
\end{landscape}

\clearpage

\begin{landscape}
  \subsubsection{Using LLL and $r = 1$}
  \label{app:rsa-ifp-lll-scaled-up-d-m}

  \begin{table}[h!]
    \begin{center}
      {\setlength{\tabcolsep}{0.45em}
      \renewcommand{\arraystretch}{1.25}
      \begin{tabular}{c|ccccc|r|r|cccc|r S[table-format=1.3] | r S[table-format=2.2]}
        \thickhline
        & \multicolumn{7}{c|}{\bfseries IFP via Regev~\cite{regev23} with~\cite{rv23v3, ragavan24, eg23-dlog}}
        & \multicolumn{8}{c}{\bfseries RSA IFP via Ekerå--Håstad~\cite{ekera-hastad, ekera-pp, ekera-short-success}} \\
        &
        &&&&& \multicolumn{1}{c|}{\bfseries per run} & \multicolumn{1}{c|}{\bfseries overall}
        &&&&& \multicolumn{2}{c|}{\bfseries per run} & \multicolumn{2}{c}{\bfseries overall}
        \\
        $\ceil{\, \log N \,}$ &
        $d$ & $m$ & $C$ & $\log D$ & $K$ & \#ops & \#ops &
        $m$ & $s$ & $\ell$ & $n$ & \#ops & \multicolumn{1}{r|}{adv} & \#ops & \multicolumn{1}{r}{adv} \\
        \thickhline
        2048 & 181 & 181 & 1.01 &  51 &  74 &  1480 &  267880 & 1023 & -- &  993 &  1 &   602 &    2.45 &    602 &  444    \\
             &     &     &      &     &     &       &         &      & 17 &   61 & 20 &   230 &    6.43 &   4600 &   58.2  \\
        \hline
        3072 & 222 & 222 & 1.01 &  61 &  88 &  1760 &  390720 & 1535 & -- & 1505 &  1 &   910 &    1.93 &    910 &  429    \\
             &     &     &      &     &     &       &         &      & 21 &   74 & 24 &   338 &    5.20 &   8112 &   48.1  \\
        \hline
        4096 & 256 & 256 & 1.01 &  70 & 101 &  2020 &  517120 & 2047 & -- & 2017 &  1 &  1218 &    1.65 &   1218 &  424    \\
             &     &     &      &     &     &       &         &      & 24 &   86 & 27 &   444 &    4.54 &  11988 &   43.1  \\
        \hline
        6144 & 314 & 314 & 1.01 &  85 & 123 &  2460 &  772440 & 3071 & -- & 3041 &  1 &  1832 &    1.34 &   1832 &  421    \\
             &     &     &      &     &     &       &         &      & 31 &  100 & 34 &   656 &    3.75 &  22304 &   34.6  \\
        \hline
        8192 & 362 & 362 & 1.01 &  97 & 140 &  2800 & 1013600 & 4095 & -- & 4065 &  1 &  2446 &    1.14 &   2446 &  414    \\
             &     &     &      &     &     &       &         &      & 34 &  121 & 37 &   868 &    3.22 &  32116 &   31.5  \\
        \thickhline
      \end{tabular}}
    \end{center}
    \caption{Comparison between Ekerå--Gärtner's extension~\cite{eg23-dlog} of Regev's algorithm~\cite{regev23} (with~\cite{rv23v3, ragavan24}, LLL, $r = 1$, optimal~$d$ and~$m$) and Ekerå--Håstad's variation~\cite{ekera-hastad, ekera-pp, ekera-short-success} of Shor's algorithm~\cite{shor94, shor97} (with $\cexp = 10$) for the RSA IFP.
    For further information on this table and how to interpret it, see Sect.~\ref{sect:rsa-ifp-lll-scaled-up-d-m} and App.~\ref{app:tables}.}
    \label{table:rsa-ifp-lll-scaled-up-d-m}
  \end{table}
\end{landscape}

\clearpage

\begin{landscape}
  \subsubsection{Using BKZ-200 and $r = 1$}
  \label{app:rsa-ifp-bkz-200-scaled-up-d-m}

  \begin{table}[h!]
    \begin{center}
      {\setlength{\tabcolsep}{0.45em}
      \renewcommand{\arraystretch}{1.25}
      \begin{tabular}{c|ccccc|r|r|cccc|r S[table-format=1.2] | r S[table-format=3.1]}
        \thickhline
        & \multicolumn{7}{c|}{\bfseries IFP via Regev~\cite{regev23} with~\cite{rv23v3, ragavan24, eg23-dlog}}
        & \multicolumn{8}{c}{\bfseries RSA IFP via Ekerå--Håstad~\cite{ekera-hastad, ekera-pp, ekera-short-success}} \\
        &
        &&&&& \multicolumn{1}{c|}{\bfseries per run} & \multicolumn{1}{c|}{\bfseries overall}
        &&&&& \multicolumn{2}{c|}{\bfseries per run} & \multicolumn{2}{c}{\bfseries overall}
        \\
        $\ceil{\, \log N \,}$ &
        $d$ & $m$ & $C$ & $\log D$ & $K$ & \#ops & \#ops &
        $m$ & $s$ & $\ell$ & $n$ & \#ops & \multicolumn{1}{r|}{adv} & \#ops & \multicolumn{1}{r}{adv} \\
        \thickhline
        2048 & 233 & 342 & 0.55 &  30 &  43 &   860 &  294120 & 1023 & -- &  993 &  1 &   602 &    1.42 &    602 &  488    \\
             &     &     &      &     &     &       &         &      & 17 &   61 & 20 &   230 &    3.73 &   4600 &   63.9  \\
        \hline
        3072 & 327 & 419 & 0.54 &  36 &  52 &  1040 &  435760 & 1535 & -- & 1505 &  1 &   910 &    1.14 &    910 &  478    \\
             &     &     &      &     &     &       &         &      & 21 &   74 & 24 &   338 &    3.07 &   8112 &   53.7  \\
        \hline
        4096 & 418 & 483 & 0.54 &  40 &  58 &  1160 &  560280 & 2047 & -- & 2017 &  1 &  1218 &    0.95 &   1218 &  460    \\
             &     &     &      &     &     &       &         &      & 24 &   86 & 27 &   444 &    2.61 &  11988 &   46.7  \\
        \hline
        6144 & 591 & 592 & 0.53 &  48 &  69 &  1380 &  816960 & 3071 & -- & 3041 &  1 &  1832 &    0.75 &   1832 &  445    \\
             &     &     &      &     &     &       &         &      & 31 &  100 & 34 &   656 &    2.10 &  22304 &   36.6  \\
        \hline
        8192 & 683 & 683 & 0.53 &  54 &  78 &  1560 & 1065480 & 4095 & -- & 4065 &  1 &  2446 &    0.64 &   2446 &  435    \\
             &     &     &      &     &     &       &         &      & 34 &  121 & 37 &   868 &    1.79 &  32116 &   33.1  \\
        \thickhline
      \end{tabular}}
    \end{center}
    \caption{Comparison between Ekerå--Gärtner's extension~\cite{eg23-dlog} of Regev's algorithm~\cite{regev23} (with~\cite{rv23v3, ragavan24}, BKZ-200, $r = 1$, optimal~$d$ and~$m$) and Ekerå--Håstad's variation~\cite{ekera-hastad, ekera-pp, ekera-short-success} of Shor's algorithm~\cite{shor94, shor97} (with $\cexp = 10$) for the RSA IFP.
    For further information on this table and how to interpret it, see Sect.~\ref{sect:rsa-ifp-bkz-200-scaled-up-d-m} and App.~\ref{app:tables}.}
    \label{table:rsa-ifp-bkz-200-scaled-up-d-m}
  \end{table}
\end{landscape}

\clearpage

\begin{landscape}
  \subsubsection{Using LLL and optimal~$r$}
  \label{app:rsa-ifp-lll-optimal-r}

  \begin{table}[h!]
    \begin{center}
      {\setlength{\tabcolsep}{0.45em}
      \renewcommand{\arraystretch}{1.25}
      \begin{tabular}{c|cccccc|r|r|cccc|r S[table-format=1.2] | r S[table-format=3.1]}
        \thickhline
        & \multicolumn{8}{c|}{\bfseries IFP via Regev~\cite{regev23} with~\cite{rv23v3, ragavan24, eg23-dlog}}
        & \multicolumn{8}{c}{\bfseries RSA IFP via Ekerå--Håstad~\cite{ekera-hastad, ekera-pp, ekera-short-success}} \\
        &
        &&&&&& \multicolumn{1}{c|}{\bfseries per run} & \multicolumn{1}{c|}{\bfseries overall}
        &&&&& \multicolumn{2}{c|}{\bfseries per run} & \multicolumn{2}{c}{\bfseries overall}
        \\
        $\ceil{\, \log N \,}$ &
        $d$ & $m$ & $C$ & $\log D$ & $K^{(r)}$ & $r$ & \#ops & \#ops &
        $m$ & $s$ & $\ell$ & $n$ & \#ops & \multicolumn{1}{r|}{adv} & \#ops & \multicolumn{1}{r}{adv} \\
        \thickhline
        2048 &  75 & 181 & 1.21 &  59 &  29 &   4 &   986 &  178466 & 1023 & -- &  993 &  1 &   602 &    1.63 &    602 &  296    \\
             &     &     &      &     &     &     &       &         &      & 17 &   61 & 20 &   230 &    4.28 &   4600 &   38.7  \\
        \hline
        3072 & 104 & 222 & 1.16 &  69 &  34 &   4 &  1156 &  256632 & 1535 & -- & 1505 &  1 &   910 &    1.27 &    910 &  282    \\
             &     &     &      &     &     &     &       &         &      & 21 &   74 & 24 &   338 &    3.42 &   8112 &   31.6  \\
        \hline
        4096 & 131 & 256 & 1.12 &  77 &  37 &   4 &  1258 &  322048 & 2047 & -- & 2017 &  1 &  1218 &    1.03 &   1218 &  264    \\
             &     &     &      &     &     &     &       &         &      & 24 &   86 & 27 &   444 &    2.83 &  11988 &   26.8  \\
        \hline
        6144 & 183 & 314 & 1.08 &  90 &  44 &   4 &  1496 &  469744 & 3071 & -- & 3041 &  1 &  1832 &    0.82 &   1832 &  256    \\
             &     &     &      &     &     &     &       &         &      & 31 &  100 & 34 &   656 &    2.28 &  22304 &   21.0  \\
        \hline
        8192 & 233 & 362 & 1.05 & 100 &  48 &   4 &  1632 &  590784 & 4095 & -- & 4065 &  1 &  2446 &    0.67 &   2446 &  241    \\
             &     &     &      &     &     &     &       &         &      & 34 &  121 & 37 &   868 &    1.88 &  32116 &   18.3  \\
        \thickhline
      \end{tabular}}
    \end{center}
    \caption{Comparison between Ekerå--Gärtner's extension~\cite{eg23-dlog} of Regev's algorithm~\cite{regev23} (with~\cite{rv23v3, ragavan24}, LLL, optimal~$r$, optimal~$d$ and~$m$) and Ekerå--Håstad's variation~\cite{ekera-hastad, ekera-pp, ekera-short-success} of Shor's algorithm~\cite{shor94, shor97} (with $\cexp = 10$) for the RSA IFP.
    For further information on this table and how to interpret it, see Sect.~\ref{sect:rsa-ifp-lll-optimal-r} and App.~\ref{app:tables}.}
    \label{table:rsa-ifp-lll-optimal-r}
  \end{table}
\end{landscape}

\clearpage

\begin{landscape}
  \subsubsection{Using BKZ-200 and optimal~$r$}
  \label{app:rsa-ifp-bkz-200-optimal-r}

  \begin{table}[h!]
    \begin{center}
      {\setlength{\tabcolsep}{0.45em}
      \renewcommand{\arraystretch}{1.25}
      \begin{tabular}{c|cccccc|r|r|cccc|r S[table-format=1.2] | r S[table-format=3.1]}
        \thickhline
        & \multicolumn{8}{c|}{\bfseries IFP via Regev~\cite{regev23} with~\cite{rv23v3, ragavan24, eg23-dlog}}
        & \multicolumn{8}{c}{\bfseries RSA IFP via Ekerå--Håstad~\cite{ekera-hastad, ekera-pp, ekera-short-success}} \\
        &
        &&&&&& \multicolumn{1}{c|}{\bfseries per run} & \multicolumn{1}{c|}{\bfseries overall}
        &&&&& \multicolumn{2}{c|}{\bfseries per run} & \multicolumn{2}{c}{\bfseries overall}
        \\
        $\ceil{\, \log N \,}$ &
        $d$ & $m$ & $C$ & $\log D$ & $K^{(r)}$ & $r$ & \#ops & \#ops &
        $m$ & $s$ & $\ell$ & $n$ & \#ops & \multicolumn{1}{r|}{adv} & \#ops & \multicolumn{1}{r}{adv} \\
        \thickhline
        2048 & 131 & 342 & 0.67 &  35 &  28 &   2 &   728 &  248976 & 1023 & -- &  993 &  1 &   602 &    1.20 &    602 &  413    \\
             &     &     &      &     &     &     &       &         &      & 17 &   61 & 20 &   230 &    3.16 &   4600 &   54.1  \\
        \hline
        3072 & 183 & 419 & 0.63 &  40 &  32 &   2 &   832 &  348608 & 1535 & -- & 1505 &  1 &   910 &    0.91 &    910 &  383    \\
             &     &     &      &     &     &     &       &         &      & 21 &   74 & 24 &   338 &    2.46 &   8112 &   42.9  \\
        \hline
        4096 & 233 & 483 & 0.61 &  44 &  35 &   2 &   910 &  439530 & 2047 & -- & 2017 &  1 &  1218 &    0.75 &   1218 &  360    \\
             &     &     &      &     &     &     &       &         &      & 24 &   86 & 27 &   444 &    2.04 &  11988 &   36.6  \\
        \hline
        6144 & 327 & 592 & 0.58 &  51 &  40 &   2 &  1040 &  615680 & 3071 & -- & 3041 &  1 &  1832 &    0.57 &   1832 &  336    \\
             &     &     &      &     &     &     &       &         &      & 31 &  100 & 34 &   656 &    1.58 &  22304 &   27.6  \\
        \hline
        8192 & 233 & 683 & 0.70 &  69 &  34 &   4 &  1156 &  789548 & 4095 & -- & 4065 &  1 &  2446 &    0.47 &   2446 &  322    \\
             &     &     &      &     &     &     &       &         &      & 34 &  121 & 37 &   868 &    1.33 &  32116 &   24.5  \\
        \thickhline
      \end{tabular}}
    \end{center}
    \caption{Comparison between Ekerå--Gärtner's extension~\cite{eg23-dlog} of Regev's algorithm~\cite{regev23} (with~\cite{rv23v3, ragavan24}, BKZ-200, optimal~$r$, optimal~$d$ and~$m$) and Ekerå--Håstad's variation~\cite{ekera-hastad, ekera-pp, ekera-short-success} of Shor's algorithm~\cite{shor94, shor97} (with $\cexp = 10$) for the RSA IFP.
    For further information on this table and how to interpret it, see Sect.~\ref{sect:rsa-ifp-bkz-200-optimal-r} and App.~\ref{app:tables}.}
    \label{table:rsa-ifp-bkz-200-optimal-r}
  \end{table}
\end{landscape}

\clearpage

\begin{landscape}
  \subsubsection{Using perfect reduction and optimal~$r$}
  \label{app:rsa-ifp-perfect-reduction}

  \begin{table}[h!]
    \begin{center}
      {\setlength{\tabcolsep}{0.45em}
      \renewcommand{\arraystretch}{1.25}
      \begin{tabular}{c|cccc|r |cccc|r S[table-format=1.2] }
        \thickhline
        & \multicolumn{5}{c|}{\bfseries IFP via Regev~\cite{regev23} with~\cite{rv23v3, ragavan24, eg23-dlog}}
        & \multicolumn{6}{c}{\bfseries RSA IFP via Ekerå--Håstad~\cite{ekera-hastad, ekera-pp}} \\
        &
        &&&& \multicolumn{1}{c|}{\bfseries per run}
        &&&&& \multicolumn{2}{c}{\bfseries per run}
        \\
        $\ceil{\, \log N \,}$ &
        $d$ & $C$ & $\log D$ & $K$ & \#ops &
        $m$ & $s$ & $\ell$ & $n$ & \#ops & \multicolumn{1}{c}{adv} \\
        \thickhline
        2048  &  131 & 0.346 &  21 &  30 &  600  &  1023 & 17 &   61 & 20 &   230 &    2.60 \\
              &  233 & 0.195 &  14 &  20 &  400  &       &    &      &    &       &    1.73 \\
        \hline
        3072  &  183 & 0.303 &  22 &  32 &  640  &  1535 & 21 &   74 & 24 &   338 &    1.89 \\
              &  327 & 0.170 &  15 &  22 &  440  &       &    &      &    &       &    1.30 \\
        \hline
        4096  &  233 & 0.275 &  23 &  33 &  660  &  2047 & 24 &   86 & 27 &   444 &    1.48 \\
              &  418 & 0.154 &  16 &  23 &  460  &       &    &      &    &       &    1.03 \\
        \hline
        6144  &  327 & 0.240 &  24 &  35 &  700  &  3071 & 31 &  100 & 34 &   656 &    1.06 \\
              &  591 & 0.133 &  17 &  25 &  500  &       &    &      &    &       &    0.76 \\
        \hline
        8192  &  418 & 0.217 &  25 &  36 &  720  &  4095 & 34 &  121 & 37 &   868 &    0.83 \\
              &  758 & 0.120 &  17 &  25 &  500  &       &    &      &    &       &    0.58 \\
        \thickhline
      \end{tabular}}
    \end{center}
    \caption{Comparison between Regev's algorithm~\cite{regev23} (with~\cite{rv23v3, ragavan24}, perfect reduction, optimal~$r = 1$, optimal~$d$ and $m \rightarrow \infty$) and Ekerå--Håstad's variation~\cite{ekera-hastad, ekera-pp, ekera-short-success} of Shor's algorithm~\cite{shor94, shor97} (with $\cexp = 10$) for the RSA IFP.
    Note that this table is special in that it considers Regev's algorithm both with and without the extensions of Ekerå and Gärtner~\cite{eg23-dlog}, and in that the number of runs $m \rightarrow \infty$ for Regev giving Ekerå--Håstad an infinite overall advantage.
    Note furthermore that since~$C$ is quite small in this table, we exceptionally tabulate~$C$ with precision~$0.001$.
    For further information on this table and how to interpret it, see Sect.~\ref{sect:rsa-ifp-perfect-reduction} and App.~\ref{app:tables}.}
    \label{table:rsa-ifp-perfect-reduction}
  \end{table}
\end{landscape}

\clearpage

\begin{landscape}
  \subsection{DLP in finite fields}

  \subsubsection{General DLP in safe-prime groups}
  \label{app:dlp-safe-prime-group-general}

  \begin{table}[h!]
    \begin{center}
      {
      \setlength{\tabcolsep}{0.45em}
      \renewcommand{\arraystretch}{1.25}
      \begin{tabular}{cc|cccccc|r|r|ccccc|r S[table-format=1.3] | r S[table-format=2.2]}
        \thickhline
        && \multicolumn{8}{c|}{\bfseries DLP via Ekerå--Gärtner~\cite{eg23-dlog, regev23} with~\cite{rv23v3, ragavan24}}
        & \multicolumn{9}{c}{\bfseries DLP via Ekerå's variation~\cite{ekera-revisiting} of Shor~\cite{shor94, shor97}} \\
        &&
        &&&&&& \multicolumn{1}{c|}{\bfseries per run} & \multicolumn{1}{c|}{\bfseries overall}
        &&&&&& \multicolumn{2}{c|}{\bfseries per run} & \multicolumn{2}{c}{\bfseries overall} \\
          $\ceil{\, \log N \,}$ & $z$ & $d$ & $m$ & $C$ & $\log D$ & $K^{(r)}$ & $r$ & \#ops & \#ops & $m$ & $s$ & $\varsigma$ & $\ell$ & $n$ & \#ops & \multicolumn{1}{c|}{adv} & \#ops & \multicolumn{1}{c}{adv} \\
        \thickhline
        2048 & 112 & 131 & 342 & 0.67 &  35 &  28 &   2 &   736 &  251712 & 2047 &  1 &  0 & 2047 &  1 &   820 &    0.90 &    820 &  306    \\
             &     &     &     &      &     &     &     &       &         &      & 24 & 11 &   86 & 27 &   430 &    1.71 &  11610 &   21.6  \\
        \hline
        3072 & 128 & 183 & 419 & 0.63 &  40 &  32 &   2 &   840 &  351960 & 3071 &  1 &  0 & 3071 &  1 &  1230 &    0.68 &   1230 &  286    \\
             &     &     &     &      &     &     &     &       &         &      & 31 & 12 &  100 & 34 &   638 &    1.31 &  21692 &   16.2  \\
        \hline
        4096 & 152 & 233 & 483 & 0.61 &  44 &  35 &   2 &   920 &  444360 & 4095 &  1 &  0 & 4095 &  1 &  1638 &    0.56 &   1638 &  271    \\
             &     &     &     &      &     &     &     &       &         &      & 34 & 12 &  121 & 37 &   846 &    1.08 &  31302 &   14.1  \\
        \hline
        6144 & 176 & 327 & 592 & 0.58 &  51 &  40 &   2 &  1052 &  622784 & 6143 &  1 &  0 & 6143 &  1 &  2458 &    0.43 &   2458 &  253    \\
             &     &     &     &      &     &     &     &       &         &      & 37 & 12 &  167 & 40 &  1266 &    0.83 &  50640 &   12.2  \\
        \hline
        8192 & 200 & 233 & 683 & 0.70 &  69 &  34 &   4 &  1170 &  799110 & 8191 &  1 &  0 & 8191 &  1 &  3278 &    0.36 &   3278 &  243    \\
             &     &     &     &      &     &     &     &       &         &      & 40 & 12 &  205 & 43 &  1682 &    0.69 &  72326 &   11.0  \\
        \thickhline
      \end{tabular}}
    \end{center}
    \caption{Comparison between Ekerå--Gärtner's extension~\cite{eg23-dlog} of Regev's algorithm~\cite{regev23} (with~\cite{rv23v3, ragavan24}, BKZ-200, optimal~$r$, optimal~$d$ and~$m$) and Ekerå's variation~\cite{ekera-revisiting} of Shor's algorithm~\cite{shor94, shor97} (with $\cexp = 10$) for the general DLP in safe-prime groups.
    For further information on this table and how to interpret it, see Sect.~\ref{sect:dlp-safe-prime-group-general} and App.~\ref{app:tables}.}
    \label{table:dlp-safe-prime-group-general}
  \end{table}
\end{landscape}

\clearpage

\begin{landscape}
  \subsubsection{Short DLP in safe-prime groups}
  \label{app:dlp-safe-prime-group-short}

  \begin{table}[h!]
    \begin{center}
      {\setlength{\tabcolsep}{0.45em}
      \renewcommand{\arraystretch}{1.25}
      \begin{tabular}{cc|cccccc|r|r|cccc|r S[table-format=2.2] | r S[table-format=3.1]}
        \thickhline
        && \multicolumn{8}{c|}{\bfseries DLP via Ekerå--Gärtner~\cite{eg23-dlog, regev23} with~\cite{rv23v3, ragavan24}}
        & \multicolumn{8}{c}{\bfseries short DLP via Ekerå--Håstad~\cite{ekera-hastad, ekera-pp, ekera-short-success}} \\
        &&
        &&&&&& \multicolumn{1}{c|}{\bfseries per run} & \multicolumn{1}{c|}{\bfseries overall}
        &&&&& \multicolumn{2}{c|}{\bfseries per run} & \multicolumn{2}{c}{\bfseries overall} \\
        $\ceil{\, \log N \,}$ & $z$ & $d$ & $m$ & $C$ & $\log D$ & $K^{(r)}$ & $r$ & \#ops & \#ops & $m$ & $s$ & $\ell$ & $n$ & \#ops & \multicolumn{1}{c|}{adv} & \#ops & \multicolumn{1}{c}{adv} \\
        \thickhline
        2048 & 112 & 131 & 342 & 0.67 &  35 &  28 &   2 &   736 &  251712 &  224 & -- &  194 &  1 &   124 &    5.93 &    124 & 2020    \\
             &     &     &     &      &     &     &     &       &         &      &  7 &   32 & 10 &    58 &   12.6  &    580 &  433    \\
        \hline
        3072 & 128 & 183 & 419 & 0.63 &  40 &  32 &   2 &   840 &  351960 &  256 & -- &  226 &  1 &   142 &    5.91 &    142 & 2470    \\
             &     &     &     &      &     &     &     &       &         &      &  8 &   32 & 11 &    64 &   13.1  &    704 &  499    \\
        \hline
        4096 & 152 & 233 & 483 & 0.61 &  44 &  35 &   2 &   920 &  444360 &  304 & -- &  274 &  1 &   172 &    5.34 &    172 & 2580    \\
             &     &     &     &      &     &     &     &       &         &      &  9 &   34 & 12 &    76 &   12.1  &    912 &  487    \\
        \hline
        6144 & 176 & 327 & 592 & 0.58 &  51 &  40 &   2 &  1052 &  622784 &  352 & -- &  322 &  1 &   200 &    5.26 &    200 & 3110    \\
             &     &     &     &      &     &     &     &       &         &      & 10 &   36 & 13 &    86 &   12.2  &   1118 &  557    \\
        \hline
        8192 & 200 & 233 & 683 & 0.70 &  69 &  34 &   4 &  1170 &  799110 &  400 & -- &  370 &  1 &   228 &    5.13 &    228 & 3500    \\
             &     &     &     &      &     &     &     &       &         &      & 11 &   37 & 14 &    96 &   12.1  &   1344 &  594    \\
        \thickhline
      \end{tabular}}
    \end{center}
    \caption{Comparison between Ekerå--Gärtner's extension~\cite{eg23-dlog} of Regev's algorithm~\cite{regev23} (with~\cite{rv23v3, ragavan24}, BKZ-200, optimal~$r$, optimal~$d$ and~$m$) and Ekerå--Håstad's variation~\cite{ekera-hastad, ekera-pp, ekera-short-success} of Shor's algorithm~\cite{shor94, shor97} (with $\cexp = 10$) for the short DLP in safe-prime groups.
    For further information on this table and how to interpret it, see Sect.~\ref{sect:dlp-safe-prime-group-short} and App.~\ref{app:tables}.}
    \label{table:dlp-safe-prime-group-short}
  \end{table}
\end{landscape}

\clearpage

\begin{landscape}
  \subsubsection{DLP in Schnorr groups}
  \label{app:dlp-schnorr-group}

  \begin{table}[h!]
    \begin{center}
      {
      \setlength{\tabcolsep}{0.45em}
      \renewcommand{\arraystretch}{1.25}
      \begin{tabular}{cc|cccccc|r|r|ccccc|r S[table-format=2.2] | r S[table-format=3.1]}
        \thickhline
        && \multicolumn{8}{c|}{\bfseries DLP via Ekerå--Gärtner~\cite{eg23-dlog, regev23} with~\cite{rv23v3, ragavan24}}
        & \multicolumn{9}{c}{\bfseries DLP via Ekerå's variation~\cite{ekera-revisiting} of Shor~\cite{shor94, shor97}} \\
        &&
        &&&&&& \multicolumn{1}{c|}{\bfseries per run} & \multicolumn{1}{c|}{\bfseries overall}
        &&&&&& \multicolumn{2}{c|}{\bfseries per run} & \multicolumn{2}{c}{\bfseries overall} \\
        $\ceil{\, \log N \,}$ & $z$ & $d$ & $m$ & $C$ & $\log D$ & $K^{(r)}$ & $r$ & \#ops & \#ops & $m$ & $s$ & $\varsigma$ & $\ell$ & $n$ & \#ops & \multicolumn{1}{c|}{adv} & \#ops & \multicolumn{1}{c}{adv} \\
        \thickhline
        2048 & 112 & 131 & 342 & 0.67 &  35 &  28 &   2 &   736 &  251712 &  224 &  1 &  0 &  224 &  1 &    90 &    8.17 &     90 & 2790    \\
             &     &     &     &      &     &     &     &       &         &      &  7 &  9 &   32 & 10 &    54 &   13.6  &    540 &  466    \\
        \hline
        3072 & 128 & 183 & 419 & 0.63 &  40 &  32 &   2 &   840 &  351960 &  256 &  1 &  0 &  256 &  1 &   104 &    8.07 &    104 & 3380    \\
             &     &     &     &      &     &     &     &       &         &      &  8 &  9 &   32 & 11 &    60 &   14.0  &    660 &  533    \\
        \hline
        4096 & 152 & 233 & 483 & 0.61 &  44 &  35 &   2 &   920 &  444360 &  304 &  1 &  0 &  304 &  1 &   122 &    7.54 &    122 & 3640    \\
             &     &     &     &      &     &     &     &       &         &      &  9 & 10 &   34 & 12 &    70 &   13.1  &    840 &  529    \\
        \hline
        6144 & 176 & 327 & 592 & 0.58 &  51 &  40 &   2 &  1052 &  622784 &  352 &  1 &  0 &  352 &  1 &   142 &    7.40 &    142 & 4380    \\
             &     &     &     &      &     &     &     &       &         &      & 10 & 10 &   36 & 13 &    80 &   13.1  &   1040 &  598    \\
        \hline
        8192 & 200 & 233 & 683 & 0.70 &  69 &  34 &   4 &  1170 &  799110 &  400 &  1 &  0 &  400 &  1 &   160 &    7.31 &    160 & 4990    \\
             &     &     &     &      &     &     &     &       &         &      & 11 & 10 &   37 & 14 &    90 &   13.0  &   1260 &  634    \\
        \thickhline
      \end{tabular}}
    \end{center}
    \caption{Comparison between Ekerå--Gärtner's extension~\cite{eg23-dlog} of Regev's algorithm~\cite{regev23} (with~\cite{rv23v3, ragavan24}, BKZ-200, optimal~$r$, optimal~$d$ and~$m$) and Ekerå's variation~\cite{ekera-revisiting} of Shor's algorithm~\cite{shor94, shor97} (with $\cexp = 10$) for the DLP in Schnorr groups.
    For further information on this table and how to interpret it, see Sect.~\ref{sect:dlp-schnorr-group} and App.~\ref{app:tables}.}
    \label{table:dlp-schnorr-group}
  \end{table}
\end{landscape}

\end{document}